\documentclass{revtex4}
\usepackage{amsthm,amssymb,amsmath}				
\usepackage{graphicx,comment}
\usepackage{float}   
 \usepackage{xcolor}

\begin{document}

\title{Klein-Gordon particles in a quasi-pointlike global monopole spacetime and a Wu-Yang magnetic monopole: invariance and isospectrality.}
\author{Omar Mustafa}
\email{omar.mustafa@emu.edu.tr}
\affiliation{Department of Physics, Eastern Mediterranean University, G. Magusa, north
Cyprus, Mersin 10 - Turkey.}

\begin{abstract}
\textbf{Abstract:}\ We use two correlated metric functions and transform/deform the pointlike global monopole (PGM) spacetime metric into a quasi-PGM (QPGM) spacetime one. We study Klein-Gordon (KG) particles
(manifestly introduced by the non-minimal coupling form of the operator $%
\tilde{D}_{\mu }=D_{\mu }+\mathcal{F}_{\mu }$, with $\mathcal{F}_{\mu }$ $%
\in 
%TCIMACRO{\U{211d} }%
%BeginExpansion
\mathbb{R}
%EndExpansion
$, and $D_{\mu }=\partial _{\mu }-ieA_{\mu }$) in such QPGM spacetime background. We show that the KG-particles in the QPGM spacetime are isospectral and invariant with KG-particles in the PGM spacetime, provided that the two metric functions are correlated. The Wu-Yang magnetic monopole (WYMM) is also included in the process. We report the effects of both monopoles PGM/QPGM and WYMM on the spectroscopic structure of two models, KG-oscillators and KG-Coulomb particles in such spacetime background. Moreover, we discuss KG-Coulomb particles and shifted KG-oscillators in a PGM spacetime and a WYMM, where such models are manifestly introduced by some Lorentz scalar interaction potentials $S(r)$. The current study is not only of fundamental observability nature but also of pedagogical interest in quantum gravity.

\textbf{PACS }numbers\textbf{: }05.45.-a, 03.50.Kk, 03.65.-w

\textbf{Keywords:} Klein-Gordon (KG) oscillators, KG-Coulomb particles, Shifted KG-oscillators, Quasi-pointlike global monopole, Wu-Yang magnetic monopole, isospectrality and invariance.
\end{abstract}

\maketitle

\section{Introduction}

During the phase transition in the early universe, Grand Unified Theories have predicted several kinds of topological defects that are manifestly introduced by the topology of the vacuum manifold \cite{Re1,Re2,Re021}. The effects of such topological defects include (but not restricted to) modifications in the spectroscopic structures of the quantum mechanical systems and have, therefore, stimulated research attention in many areas of physics. For example, in condensed matter physics and gravitation, and within the geometrical theory of topological defects, linear defects are associated with dislocations (i.e., torsions) and disclinations (curvatures) \cite{Re0211,Re0212}. The topological defects may include cosmic string \cite%
{Re021,Re3,Re4,Re5,Re6}, domain walls \cite{Re2,Re021}, and pointlike global monopole (PGM) \cite{Re7}. Global monopoles formed as a consequence of a spontaneous global symmetry breaking (i.e., global $O(3)$ symmetry is broken to $U\left( 1\right) $) behave like elementary particles, where most of their energies are concentrated near the monopole core \cite{Re7}. Global monopoles are zero-dimensional (pointlike) objects that do not introduce gravitational interactions but they modify the geometry of spacetime \cite%
{Re3,Re6,Re7,Re8}. They are spherically symmetric topological defects that admit the general static metric%
\begin{equation}
ds^{2}=-B\left( r\right) \,dt^{2}+A\left( r\right) \,dr^{2}+r^{2}\left(
d\theta ^{2}+\sin ^{2}\theta \,d\varphi ^{2}\right) .  \label{eq1}
\end{equation}

In their study, Barriola and Vilenkin \cite{Re7} have shown that the PGM, effectively, exerts no gravitational force. The space around and outside the PGM has a solid deficit angle that deflects all light. Moreover, they have found that the correlation between $B\left( r\right) $ and $A\left( r\right) $ is given by%
\begin{equation}
B\left( r\right) =A\left( r\right) ^{-1}=1-8\pi G\eta ^{2}-\frac{2GM}{r},
\label{eq2}
\end{equation}%
where $M$ is a constant of integration and in the flat space $M\sim M_{core}$ ( $M_{core}$ is the mass of the PGM core). \ By neglecting the mass term and rescaling the variables $r$ and $t$ \cite{Re7}, one may rewrite the
PGM metric as 
\begin{equation}
ds^{2}=-dt^{2}+\frac{1}{\alpha ^{2}}dr^{2}+r^{2}\left( d\theta ^{2}+\sin
^{2}\theta \,d\varphi ^{2}\right) ,  \label{e01}
\end{equation}%
where $0<\alpha ^{2}=1-8\pi G\eta ^{2}\leq 1$, $\alpha $ is a PGM parameter that depends on the energy scale $\eta $, and $G$ is the gravitational constant \cite{Re7,Re8,Re81,Re9,Re10}. This has stimulated research attention so that PGM effects are studied, for example, within $f\left( R\right) $ theory \cite{Re81}, vacuum polarization effects in the presence of a Wu-Yang\ \ magnetic monopole (WYMM) \cite%
{Re101,Re1011,Re102}, gravitating magnetic monopole \cite{Re103}, Dirac and Klein-Gordon (KG) oscillators \cite{Re11}, Schr\"{o}dinger oscillators \cite%
{Re9}, KG particles with a dyon, magnetic flux and scalar potential \cite%
{Re8}, bosons in Aharonov-Bohm flux field and a Coulomb potential \cite%
{Re131}, Schr\"{o}dinger particles in a Kratzer potential \cite{Re132}, Schr\"{o}dinger particles in a Hulth\.{e}n potential \cite{Re1321} , scattering by a monopole \cite{Re133}, Schr\"{o}dinger particles in a Hulth\.{e}n plus Kratzer potential \cite{Re1331}, KG-oscillators and AB-effect \cite{Re1332}, to mention a few. In fact, the influence of topological defects in spacetime on the spectroscopic structure of quantum mechanical systems have been a subject of research attention over the years. The harmonic oscillator is studied in the context of Dirac and Klein-Gordon (KG) \cite%
{Re11,Re12,Re121,Re13,Re14,Re15,Re16,Re17,Re18,Re19,Re20,Re21,Re211,Re22,Re23,Re24,Re25,Re26,Re27,Re271,Re272} in different spacetime backgrounds.

On the other hand, the WYMM \cite{Re101,Re1011,Re102} is a theoretical concept that is yet to be observed as a physical particle in nature. However, it remains an interesting model for modern theoretical particle physics and topological properties in gauge theories. It would, therefore, be interesting to study KG-particles in a PGM and a quasi-point-like global monopole spacetime (QPGM) and a WYMM. QPGM spacetime is the notion we shall adopt throughout to refer to a deformed/transformed (through a point
canonical transformation) PGM spacetime. Such a study would facilitate not only exact solvability of quantum gravity models, but would also be useful for pedagogical interest in quantum gravity. This should represent the motivation of the current methodical proposal. To the best of our knowledge, no such study have ever been done before.

We organize our paper in the following manner. In section 2, we derive the KG-particles' equation in a QPGM and a 4-vector potential $A_{\mu }=\left(
0,0,A_{\varphi },0\right) $. Where we use a point canonical transformation so that the PGM metric would eventually indulge two correlated metric functions that makes the notion QPGM spacetime unavoidable in the process. We use our findings for the KG-particles in a QPGM spacetime and include the WYMM in section 3. Hereby, we show that the KG-particles in a QPGM spacetime and the WYMM are isospectral and invariant with KG-particles in the regular PGM spacetime and a WYMM. In section 4, we discuss KG-oscillators in a QPGM spacetime and a WYMM and report the effects of
both monopoles on the spectroscopic structure of the KG-oscillators. In section 5, we discuss KG-Coulomb particles in a QPGM spacetime and a WYMM, where the Coulombic nature of such KG-particles is
introduced by the non-minimal coupling form of the operator $\tilde{D}_{\mu
}=D_{\mu }+\mathcal{F}_{\mu }$, with $\mathcal{F}_{\mu }$ $\in 
%TCIMACRO{\U{211d} }%
%BeginExpansion
\mathbb{R}
%EndExpansion
$, and $D_{\mu }=\partial _{\mu }-ieA_{\mu }$ is the gauge-covariant derivative that admits minimal coupling. The KG-oscillators and KG-Coulomb particles in such spacetime background are not only of fundamental observability nature but also of pedagogical interest in quantum gravity. For both KG-oscillators and KG-Coulomb particles in a QPGM spacetime and a WYMM, we use two sets of the metric functions. Moreover, we discuss KG-Coulomb particles (in section 6) and a shifted KG-oscillators (in section 7) in a PGM spacetime and a WYMM, where the interactions' nature of which is introduced by some Lorentz scalar interaction potentials $S(r)$. We conclude in section 8.

\section{KG-particles in a QPGM spacetime background}

Consider KG-particles interacting with a PGM with a spacetime metric given by (\ref{e01}) and subjected to a point canonical transformation (PCT) in the form of 
\begin{equation}
r=\int \sqrt{f\left( \rho \right) }d\rho =\sqrt{q\left( \rho \right) }\rho
\Leftrightarrow \sqrt{f\left( \rho \right) }=\sqrt{q\left( \rho \right) }%
\left[ 1+\frac{q^{\prime }\left( \rho \right) }{2q\left( \rho \right) }\rho %
\right] .  \label{e02}
\end{equation}%
Then the PGM metric (\ref{e01}) transforms into 
\begin{equation}
ds^{2}=-dt^{2}+\frac{f\left( \rho \right) }{\alpha ^{2}}d\rho ^{2}+q\left(
\rho \right) \,\rho ^{2}\left[ d\theta ^{2}+\sin ^{2}\theta \,d\varphi ^{2}%
\right] .  \label{e03}
\end{equation}%
At this point, one should notice that $q\left( \rho \right) $ and $f\left(
\rho \right) $ are two correlated functions and are dimensionless
positive-valued scalar multipliers, where $q\left( \rho \right)
=1\Rightarrow f\left( \rho \right) =1$ recovers the PGM metric (\ref{e01}). This would not change the radial direction nor the range of the radial coordinates $\rho $ or $r$ ( i.e., $r\in \left[ 0,\infty \right]\Longrightarrow \rho \in \left[ 0,\infty \right] $). They are commonly called metric functions so that $q\left( \rho \right) ,f\left( \rho \right)
\in 
%TCIMACRO{\U{211d} }%
%BeginExpansion
\mathbb{R}
%EndExpansion
$.

Consequently, the corresponding deformed/transformed metric tensor is 
\begin{equation}
g_{\mu \nu }=\left( 
\begin{tabular}{cccc}
$-1$ & $0$ & $0$ & $0$ \\ 
$0$ & $\;f\left( \rho \right) /\alpha ^{2}$ & $0$ & $0\medskip $ \\ 
$0$ & $0$ & $q\left( \rho \right) \,\rho ^{2}$ & $0\medskip $ \\ 
$0$ & $0$ & $0$ & $q\left( \rho \right) \,\rho ^{2}\sin ^{2}\theta $%
\end{tabular}%
\right) ;\;\mu ,\nu =t,\rho ,\theta ,\varphi ,  \label{e04}
\end{equation}%
which implies 
\begin{equation*}
\det \left( g_{\mu \nu }\right) =g=-\frac{f\left( \rho \right) }{\alpha ^{2}}%
q\left( \rho \right) ^{2}\,\rho ^{4}\sin ^{2}\theta ,
\end{equation*}%
and 
\begin{equation}
g^{\mu \nu }=\left( 
\begin{tabular}{cccc}
$-1$ & $0$ & $0$ & $0$ \\ 
$0$ & $\;\alpha ^{2}/f\left( \rho \right) $ & $0$ & $0\medskip $ \\ 
$0$ & $0$ & $1/q\left( \rho \right) \,\rho ^{2}$ & $0\medskip $ \\ 
$0$ & $0$ & $0$ & $1/q\left( \rho \right) \,\rho ^{2}\sin ^{2}\theta $%
\end{tabular}%
\right) .  \label{e05}
\end{equation}%
The KG-equation then reads 
\begin{equation}
\left( \frac{1}{\sqrt{-g}}\tilde{D}_{\mu }\sqrt{-g}g^{\mu \nu }\tilde{D}%
_{\nu }+\zeta R\right) \,\Psi \left( t,\rho ,\theta ,\varphi \right) =\left(
m_{\circ }+S\left( \rho \right) \right) ^{2}\,\Psi \left( t,\rho ,\theta
,\varphi \right) ,  \label{e06}
\end{equation}%
where $\tilde{D}_{\mu }=D_{\mu }+\mathcal{F}_{\mu }$ is in a non-minimal coupling form with $\mathcal{F}_{\mu }$ $\in 
%TCIMACRO{\U{211d} }%
%BeginExpansion
\mathbb{R}
%EndExpansion
$, $D_{\mu }=\partial _{\mu }-ieA_{\mu }$ is the gauge-covariant derivative that admits minimal coupling, $A_{\nu }=\left( 0,0,A_{\varphi },0\right) $ is the 4-vector potential, $S\left( \rho \right) $ is the the Lorentz scalar potential, $m$ is the rest mass energy (i.e., $m\equiv mc^{2}$, with $\hbar
=c=1$ units to be used through out this study), $\zeta $ is an arbitrary coupling constant, and the scalar curvature $R$ reads $R=R_{\nu }^{\nu
}=2\left( 1-\alpha ^{2}\right) /q\left( \rho \right) \,\rho ^{2}$ . One should notice that the term $\mathcal{F}_{\mu }$ in the non-minimal coupling form of $\tilde{D}_{\mu }$ is used to incorporate the KG-oscillators for $%
\mathcal{F}_{\mu }=\left( 0,\mathcal{F}_{\rho },0,0\right) $, $\mathcal{F}%
_{\rho }\equiv \mathcal{F}_{\rho }\left( \rho \right) $ (c.f., e.g., \cite{Re0281,Re0282}, and the same recipe can be used for KG-Coulombic particles as well). This would, in turn, yield%
\begin{gather}
\left\{ -\partial _{t}^{2}+\frac{\alpha ^{2}}{\sqrt{f\left( \rho \right) }%
q\left( \rho \right) \rho ^{2}}\,\left( \partial _{\rho }+\mathcal{F}_{\rho
}\right) \frac{q\left( \rho \right) \rho ^{2}}{\sqrt{f\left( \rho \right) }}%
\,\left( \partial _{\rho }-\mathcal{F}_{\rho }\right) +\right. \frac{2\zeta
\left( 1-\alpha ^{2}\right) }{q\left( \rho \right) \,\rho ^{2}}+\frac{1}{%
q\left( \rho \right) \rho ^{2}}\left[ \frac{1}{\sin \theta }\,\partial
_{\theta }\,\sin \theta \,\partial _{\theta }\right.   \notag \\
\left. \left. +\frac{1}{\sin ^{2}\theta }\left( \partial _{\varphi
}-ieA_{\varphi }\right) ^{2}\right] \right\} \,\Psi \left( t,\rho ,\theta
,\varphi \right)  =\left( m_{\circ }+S\left( \rho \right) \right)
^{2}\,\Psi \left( t,\rho ,\theta ,\varphi \right) .  \label{e07}
\end{gather}%
This equation describes KG-particles in a QPGM (a deformed/transformed PGM if you wish) spacetime and a 4-vector potential $A_{\nu }=\left( 0,0,A_{\varphi },0\right)$. In the forthcoming sections, we shall consider some interesting  characterizations  of such KG-systems that are of fundamental importance in the fields of theoretical particle physics and quantum gravity.

\section{KG-particles in a QPGM spacetime and a WYMM}

Let us now consider that the KG-particles are in a PGM spacetime Eq. (\ref{e01}) and under the influence of a WYMM. Wu and Yang \cite%
{Re101} have introduced a magnetic monopole that is free of strings of singularities around it \cite{Re8,Re101,Re102}., and have defined the 4-vector potential $A_{\mu }$ in two regions, $R_{A}$ and $R_{B}$ covering the whole space, outside the magnet monopole, and overlap in $R_{AB}$ so that%
\begin{equation}
\begin{tabular}{lll}
$R_{A}:0\leq \theta <\frac{\pi }{2}+\delta \medskip ,\;$ & $\;\;r>0,\;\;$ & $%
0\leq \varphi <2\pi ,$ \\ 
$R_{B}:\frac{\pi }{2}-\delta <\theta \leq \pi ,\;$ & $\;\;r>0,\;\;$ & $0\leq
\varphi <2\pi ,\medskip $ \\ 
$R_{AB}:\frac{\pi }{2}-\delta <\theta <\frac{\pi }{2}+\delta ,$ & $\;\;r>0,\;
$ & $0\leq \varphi <2\pi ,$%
\end{tabular}
\label{e071}
\end{equation}%
where $0<\delta \leq \pi /2$. Moreover, the 4-vector potential has a non-vanishing component in each region given by%
\begin{equation}
A_{\varphi }=sg-g\cos \theta =\left\{ 
\begin{tabular}{ll}
$A_{\varphi ,A}=g\left( 1-\cos \theta \right) ,$ & for $s=1\medskip $ \\ 
$A_{\varphi ,B}=-g\left( 1+\cos \theta \right) ,$ & for $s=-1$%
\end{tabular}%
\right. ,  \label{e072}
\end{equation}%
where, $A_{\varphi ,A}$ and $A_{\varphi ,B}$ are correlated by the gauge transformation \cite{Re8,Re102} 
\begin{equation}
A_{\varphi ,A}=A_{\varphi ,B}+\frac{i}{e}S\,\partial _{\varphi
}\,S^{-1}\,;\;S=e^{2i\sigma \varphi },\;\sigma =eg.  \label{e073}
\end{equation}%
Under such settings, the substitution of $\Psi \left( t,\rho ,\theta
,\varphi \right) =e^{-iEt}\psi \left( \rho \right) Y_{\sigma ,\ell ,m}\left(
\theta ,\varphi \right) $ would allow separability of equation (\ref{e07})\
to imply 
\begin{equation}
\left\{ E^{2}+\frac{\alpha ^{2}}{\sqrt{f\left( \rho \right) }q\left( \rho
\right) \rho ^{2}}\left( \partial _{\rho }+\mathcal{F}_{\rho }\right) \,%
\frac{q\left( \rho \right) \rho ^{2}}{\sqrt{f\left( \rho \right) }}%
\,\,\left( \partial _{\rho }-\mathcal{F}_{\rho }\right) -\frac{\tilde{\lambda%
}}{q\left( \rho \right) \,\rho ^{2}}\right\} \psi \left( \rho \right)
=\left( m_{\circ }+S\left( \rho \right) \right) ^{2}\psi \left( \rho \right)
,  \label{e074}
\end{equation}%
where $\tilde{\lambda}=\lambda +2\xi \left( 1-\alpha ^{2}\right) $, and 
\begin{equation}
\left\{ \frac{1}{\sin \theta }\partial _{\theta }\sin \theta \;\partial
_{\theta }+\frac{1}{\sin ^{2}\theta }\left[ \partial _{\varphi
}-ieA_{\varphi }\right] ^{2}\right\} Y_{\sigma \ell m}\left( \theta ,\varphi
\right) =-\lambda Y_{\sigma \ell m}\left( \theta ,\varphi \right) .
\label{e075}
\end{equation}%
Obviously, we need the eigenvalues of Eq.(\ref{e075}) to solve for Eq.(\ref%
{e074}). In so doing, we use%
\begin{equation}
Y_{\sigma \ell m}\left( \theta ,\varphi \right) =e^{i\left( m+s\sigma
\right) }\ \Theta _{\sigma \ell m}\left( \theta \right) ,  \label{e076}
\end{equation}%
where $m=0,\pm 1,\pm 2,\cdots $ is the magnetic quantum number, to obtain%
\begin{equation}
\left\{ \frac{1}{\sin \theta }\partial _{\theta }\sin \theta \;\partial
_{\theta }-\frac{\left( m+\sigma \cos \theta \right) ^{2}}{\sin ^{2}\theta }%
\right\} \Theta _{\sigma \ell m}\left( \theta \right) =-\lambda \Theta
_{\sigma \ell m}\left( \theta \right) .  \label{e077}
\end{equation}%
At this point, one should notice that Eq.(\ref{e077}) is $s$-independent so that $\Theta _{\sigma \ell m}\left( \theta \right) =\left[ \Theta _{\sigma
\ell m}\left( \theta \right) \right] _{A}=\left[ \Theta _{\sigma \ell
m}\left( \theta \right) \right] _{B}$ as readily observed by Wu-Yang \cite%
{Re101}. 

Next, we use $x=\cos \theta $ and re-write Eq.(\ref{e077}) as%
\begin{equation}
\left\{ \left( 1-x^{2}\right) \,\partial _{x}^{2}-2x\,\partial _{x}-\frac{%
\left( m+\sigma \,x\right) ^{2}}{1-x^{2}}\right\} \Theta _{\sigma \ell
m}\left( \theta \right) =-\lambda \Theta _{\sigma \ell m}\left( \theta
\right) .  \label{e078}
\end{equation}%
Which, with the substitution%
\begin{equation}
\Theta _{\sigma \ell m}\left( \theta \right) =\left( 1-x\right) ^{\beta
/2}\left( 1+x\right) ^{\gamma /2}\,P_{\sigma \ell m}\left( x\right)
\,;\;\beta =|m|+\sigma ,\;\gamma =|m|-\sigma ,  \label{e079}
\end{equation}%
would yield that%
\begin{equation}
\,P_{\sigma \ell m}\left( x\right) =C\;M\left( \frac{1}{2}+|m|\pm \sqrt{%
\sigma ^{2}+\lambda +\frac{1}{4}},|m|-\sigma +1,\frac{1}{2}\left( 1+x\right)
\right) .  \label{e0791}
\end{equation}%
However, we are interested in a finite and bounded solution for quantum states. This would manifestly require that the confluent hypergeometric series $%
M\left( a,b,z\right) $ be truncated into a polynomial of order $n\geq 0$ using the condition%
\begin{equation}
\frac{1}{2}+|m|\pm \sqrt{\sigma ^{2}+\lambda +\frac{1}{4}}%
=-n\Longleftrightarrow \sigma ^{2}+\lambda =\left( n+|m|\right) \left(
n+|m|+1\right) ;\;n,|m|\,\geq 0.  \label{e0792}
\end{equation}%
One should reminded that when $\sigma =0$ (i.e., no Wu-Yang monopole effect) our $\lambda =\ell \left( \ell +1\right) $, where $\ell =0,1,2,\cdots $ is the angular momentum quantum number. Hence, without any lose of generality
of the solution, one may safely consider that $\ell =n+|m|$ and take $%
\lambda =\ell \left( \ell +1\right) -\sigma ^{2}$.  The validity of such assumptions is motivated by the fact that for $\sigma =0$ and $\alpha=1$ (i.e., flat Minkowski spacetime) one should retrieve the textbook value $\lambda =\ell \left( \ell +1\right)$. Consequently, our $\tilde{%
\lambda}$ in Eq.(\ref{e074}) now reads%
\begin{equation}
\tilde{\lambda}=\ell \left( \ell +1\right) +2\zeta \left( 1-\alpha
^{2}\right) -\sigma ^{2}=L\left( L+1\right) ,  \label{e0793}
\end{equation}%
where $L$ may very well be denoted as an irrational angular momentum quantum number. Moreover,
the so called Wu and Yang \cite{Re8,Re101} monopole harmonics $Y_{\sigma
\ell m}\left( \theta ,\varphi \right) $ are now given by%
\begin{equation}
Y_{\sigma \ell m}\left( \theta ,\varphi \right) =\left\{ 
\begin{tabular}{ll}
$e^{i\left( m+\sigma \right) \varphi }\,\left( 1-x\right) ^{\beta /2}\left(
1+x\right) ^{\gamma /2}\,P_{\sigma \ell m}\left( x\right) ;\;\medskip $ & in
region $R_{A}$ \\ 
$e^{i\left( m-\sigma \right) \varphi }\,\left( 1-x\right) ^{\beta /2}\left(
1+x\right) ^{\gamma /2}\,P_{\sigma \ell m}\left( x\right) ;\medskip $ & in
region $R_{B}$%
\end{tabular}%
\right. .  \label{0794}
\end{equation}%
As a result, our equation (\ref{e074}) reduces, with $\mathcal{F}_{\rho
}\equiv \mathcal{F}_{\rho }\left( \rho \right) $, to%
\begin{equation}
\left\{ E^{2}+\frac{\alpha ^{2}}{\sqrt{f\left( \rho \right) }q\left( \rho
\right) \rho ^{2}}\,\left( \partial _{\rho }+\mathcal{F}_{\rho }\left( \rho
\right) \right) \,\frac{q\left( \rho \right) \rho ^{2}}{\sqrt{f\left( \rho
\right) }}\,\,\left( \partial _{\rho }-\mathcal{F}_{\rho }\left( \rho
\right) \right) -\frac{L\left( L+1\right) }{q\left( \rho \right) \,\rho ^{2}}%
\right\} \psi \left( \rho \right) =\left( m_{\circ }+S\left( \rho \right)
\right) ^{2}\psi \left( \rho \right) .  \label{e0795}
\end{equation}%
Let us now oversimplify this equation using our PCT in Eq. (\ref{e02}) so that $\partial _{r}=\partial /\partial r=\partial \rho /\partial r\,\
\partial _{\rho }=\partial _{\rho }/\sqrt{f\left( \rho \right) }$ and $r=%
\sqrt{q\left( \rho \right) }\rho $ to obtain%
\begin{equation}
\left\{ \tilde{E}^{2}+\frac{1}{r^{2}}\left[ \partial _{r}+\mathcal{F}%
_{r}\left( r\right) \right] \,r^{2}\left[ \partial _{r}-\mathcal{F}%
_{r}\left( r\right) \right] -\frac{\tilde{L}\left( \tilde{L}+1\right) }{r^{2}%
}-\frac{1}{\alpha ^{2}}\left( m_{\circ }+S\left( r(\rho )\right) \right)
^{2}\right\} \psi \left( r(\rho )\right) =0,  \label{e0796}
\end{equation}%
where $\mathcal{F}_{r}\left( r\right) =$ $\mathcal{F}_{\rho }\left( \rho
\right) /\sqrt{f\left( \rho \right) }$ is used. This equation (\ref{e0796}) would, with $\psi \left( r\right) =R\left( r\right) /r$, \ read%
\begin{equation}
\left\{ \partial _{r}^{2}-\frac{\tilde{L}\left( \tilde{L}+1\right) }{r^{2}}-%
\left[ \frac{2}{r}\mathcal{F}_{r}\left( r\right) +\mathcal{F}_{r}^{^{\prime
}}\left( r\right) +\mathcal{F}_{r}\left( r\right) ^{2}\right] -2\tilde{m}%
\frac{S\left( r\right) }{\alpha }-\frac{S\left( r\right) }{\alpha ^{2}}%
^{2}+\Lambda \right\} R\left( r\right) =0,  \label{e08}
\end{equation}%
where 
\begin{equation}
\Lambda =\tilde{E}^{2}-\tilde{m}^{2},\;\tilde{E}=\frac{E}{\alpha },\;\tilde{m%
}=\frac{m_{\circ }}{\alpha },  \label{81}
\end{equation}%
and $\tilde{L}\left( \tilde{L}+1\right) =L\left( L+1\right) /\alpha ^{2}$ to imply 
\begin{equation}
\tilde{L}=-\frac{1}{2}+\frac{\sqrt{\alpha ^{2}+4L\left( L+1\right) }}{%
2\alpha }=-\frac{1}{2}+\frac{\sqrt{\alpha ^{2}+4\left[ \ell \left( \ell
+1\right) +2\zeta \left( 1-\alpha ^{2}\right) -\sigma ^{2}\right] }}{2\alpha 
}.  \label{e081}
\end{equation}%
Obviously, the result in Eq.(\ref{e081}) converges to $\tilde{L}=\ell $ \
for $\alpha =1$ and $\sigma =0$ (i.e., no effects of the PGM or the WYMM) as should be. More interestingly, our KG-particles in a QPGM spacetime, Eq, (\ref{e0795}), are isospectral and invariant with those in a PGM spacetime, Eq.(\ref{e0796}), under our correlated metric functions  (through the PCT of Eq.(\ref{e02})). Yet, in a straightforward manner, one may show that Eq.(\ref{e0796}) is nothing but the KG-particles in the regular PGM spacetime (\ref{e01}). In the following section, we shall consider some illustrative examples on such systems.

\section{KG-oscillators in a QPGM/PGM spacetime and a WYMM at $S(r)=0$ and  $\mathcal {F}(r)=\eta r$}

In this section, we shall consider KG-oscillators in a QPGM spacetime under the influence of a WYMM at $S\left( r\right) =0$. In this case, the substitution of $\mathcal{F}_{\rho }\equiv \mathcal{F}_{\rho
}\left( \rho \right) =\eta \sqrt{f\left( \rho \right) q\left( \rho \right) }%
\rho $ in Eq. (\ref{e0795}) would yield $\mathcal{F}_{r}\left( r\right) =$ $%
\mathcal{F}_{\rho }\left( \rho \right) /\sqrt{f\left( \rho \right) }=\eta \,r
$ in Eq. (\ref{e0796}). Consequently, the KG-oscillators in a QPGM spacetime
and under the influence of a WYMM are obtained so that the KG-equation (\ref{e08}), with $S\left( r\right) =0$, now reads%
\begin{equation}
\left\{ \partial _{r}^{2}-\frac{\tilde{L}\left( \tilde{L}+1\right) }{r^{2}}%
-\eta ^{2}r^{2}+\tilde{\Lambda}\right\} R\left( r\right) =0,\;\tilde{\Lambda}%
=\Lambda -3\eta .  \label{9}
\end{equation}%
Our KG-oscillators in a QPGM spacetime and a WYMM, Eq. (\ref{9}), admit exact textbook solution in the form of confluent hypergeometric series so that%
\begin{equation}
R\left( r\right) =C\,e^{-\eta r^{2}/2}\,r^{\tilde{L}+1}\,M\left( \frac{3}{4}+%
\frac{\tilde{L}}{2}-\frac{\tilde{\Lambda}}{4\eta },\tilde{L}+\frac{3}{2}%
,\eta r^{2}\right) .  \label{91}
\end{equation}%
Nevertheless, the radial function $R\left( r\right) $ should be finite and square integrable in order to be an admissible solution for a quantum mechanical system. This can be achieved by using the condition that the confluent hypergeometric series $M\left( a,b,z\right) $ is truncated into a polynomial of order $n_{r}=0,1,2,\cdots $ so that%
\begin{equation}
\frac{3}{4}+\frac{\tilde{L}}{2}-\frac{\tilde{\Lambda}}{4\eta }%
=-n_{r}\Longleftrightarrow \tilde{\Lambda}=2\eta \left( 2n_{r}+\tilde{L}+%
\frac{3}{2}\right) .  \label{92}
\end{equation}%
In a straightforward manner, one may use Eq.(\ref{81}) to obtain the energies of the KG-oscillators in a QPGM spacetime and a WYMM as%
\begin{equation}
E=\pm \sqrt{m_{\circ }^{2}+4\eta \alpha ^{2}\left( n_{r}+\frac{1}{4\alpha }%
\sqrt{\alpha ^{2}+4\ell \left( \ell +1\right) +8\zeta \left( 1-\alpha
^{2}\right) -\sigma ^{2}}+\frac{5}{4}\right) }.  \label{93}
\end{equation}%
This result is in exact accord with that reported in Eq. (38) of Bragan\c{c}a et al. \cite{Re11} when the WYMM is switched off (i.e., $\sigma =0$) and with our $\eta =m_{\circ }\omega $ of Bragan\c{c}a et al. \cite{Re11}.
\begin{figure}[!ht]  
\centering
\includegraphics[width=0.3\textwidth]{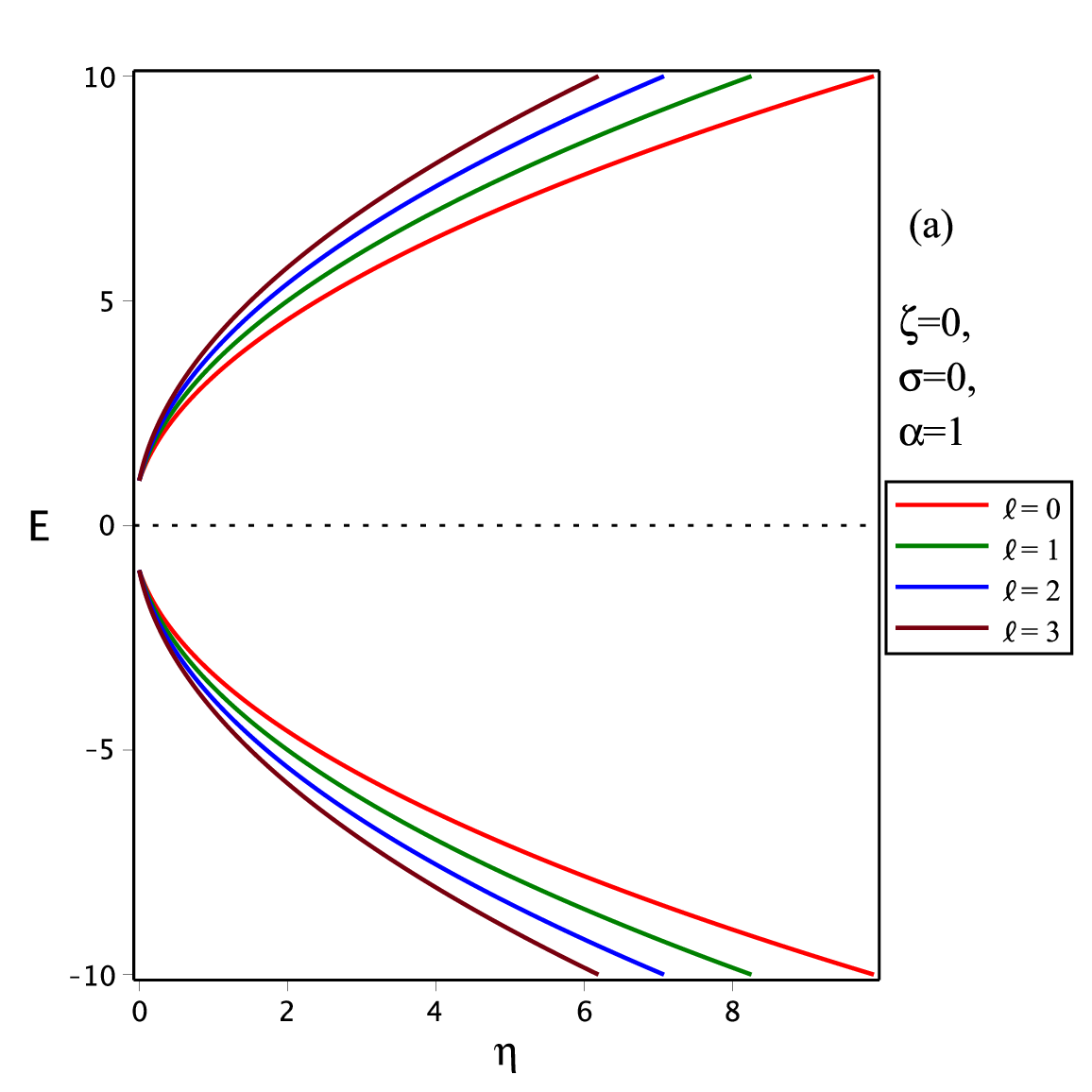}
\includegraphics[width=0.3\textwidth]{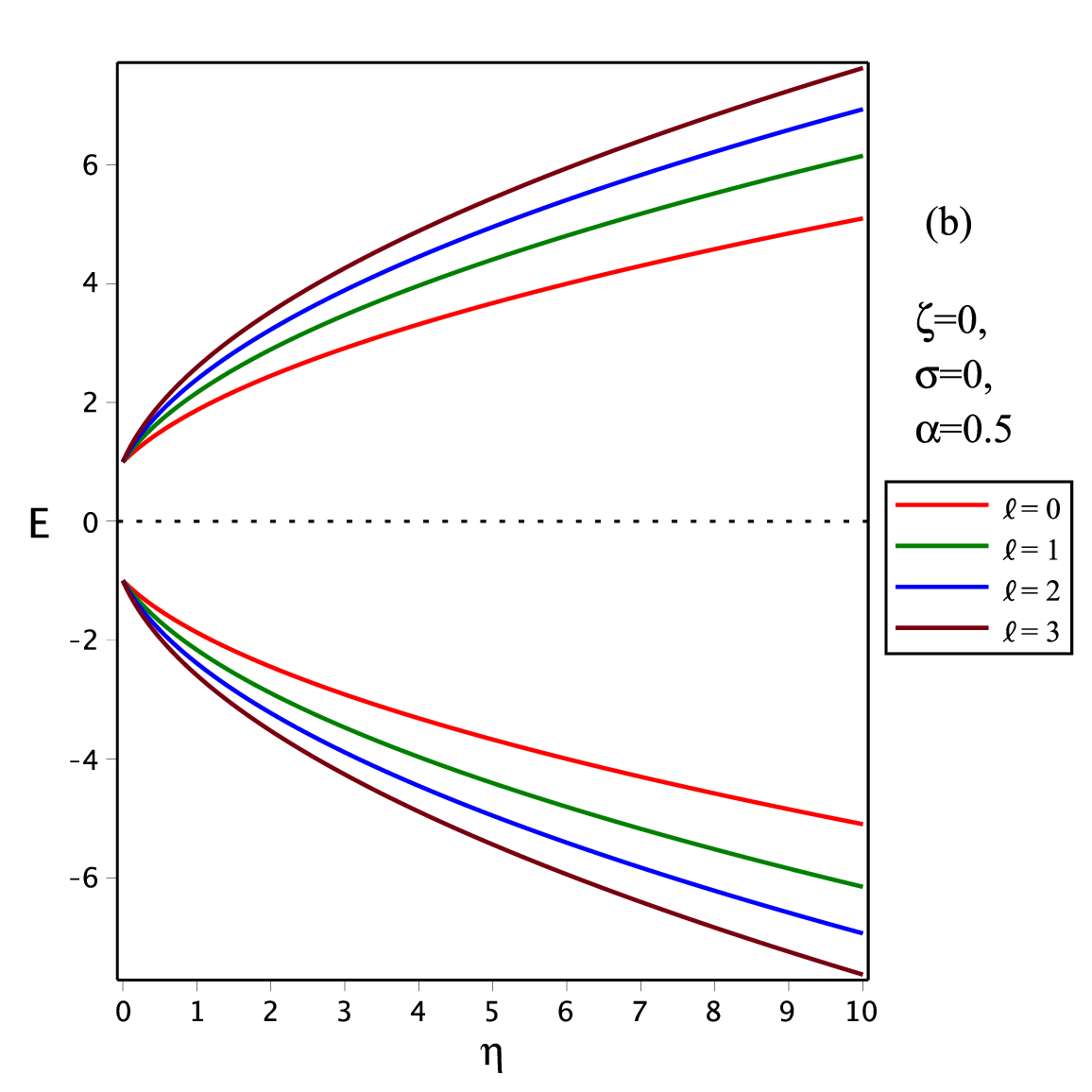} 
\includegraphics[width=0.3\textwidth]{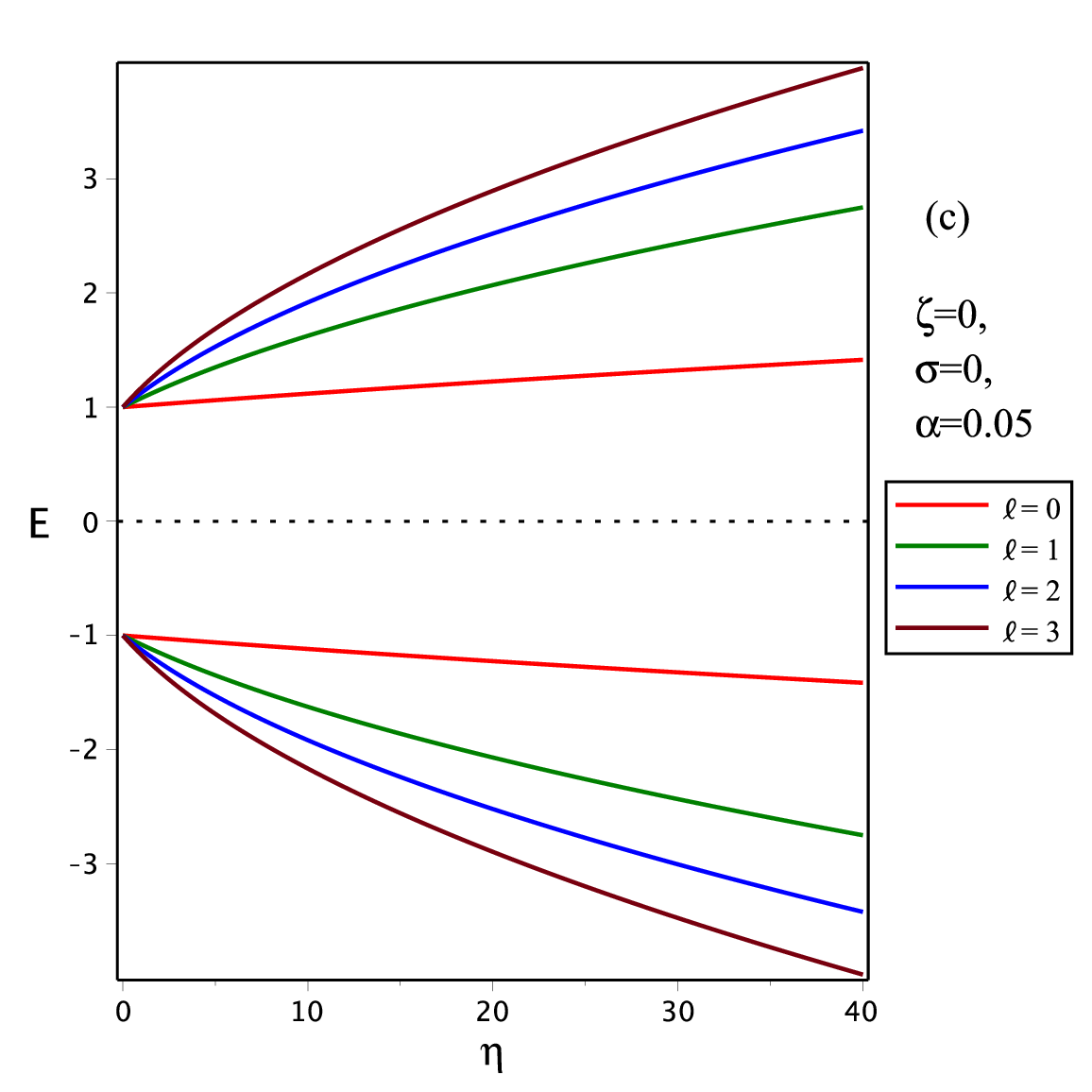}
\caption{\small 
{ The energy levels Eq.(\ref{93}) of the KG-oscillators in a
PGM/QPGM spacetime and a Wu-Yang magnetic monopole for $n_{r}=1$, $\ell
=0,1,2,3$, (a) for $\alpha =1$, (b) for $\alpha =0.5$, and\ (c) for $\alpha =0.05$.}}
\label{fig1}
\end{figure}%
\begin{figure}[!ht]  
\centering
\includegraphics[width=0.3\textwidth]{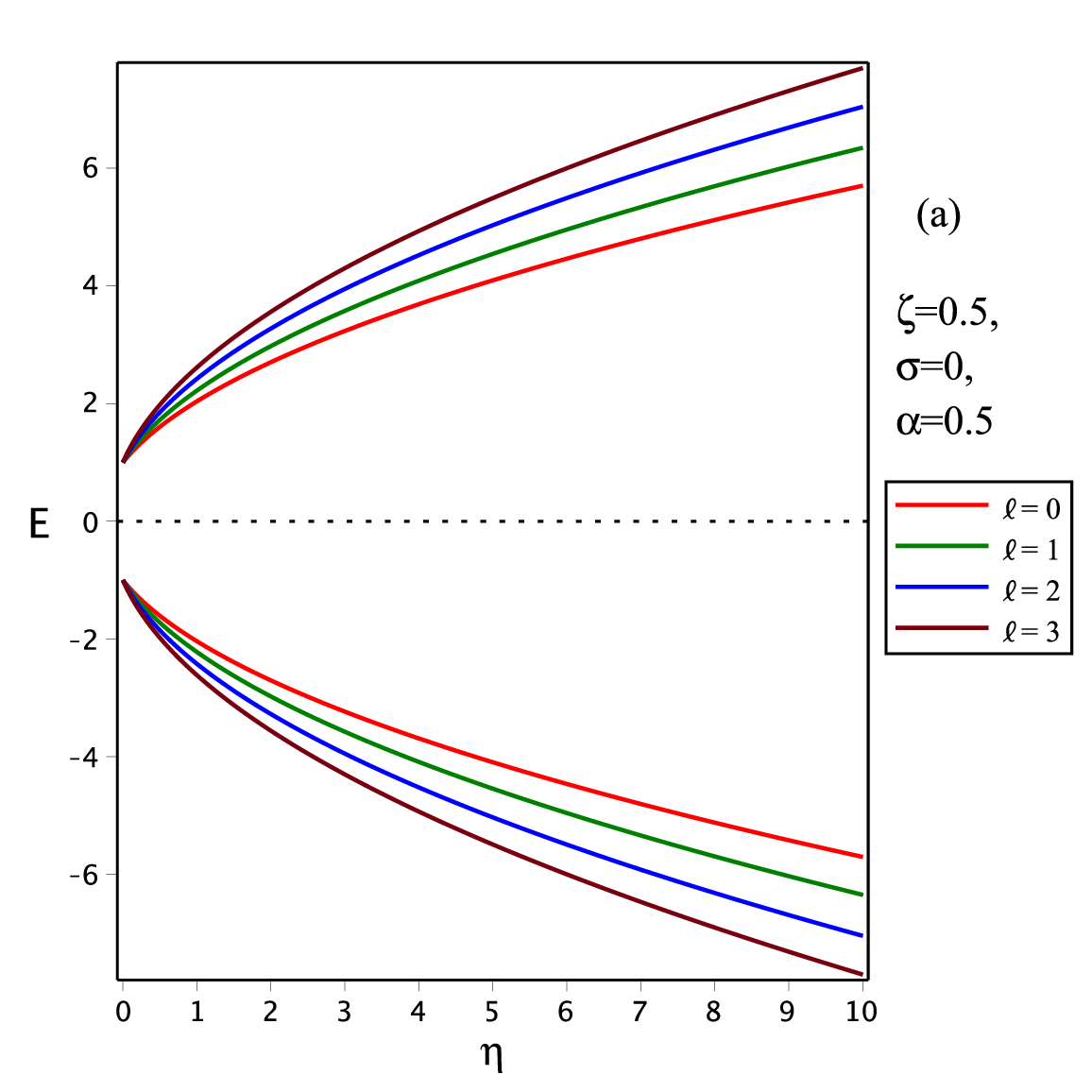}
\includegraphics[width=0.3\textwidth]{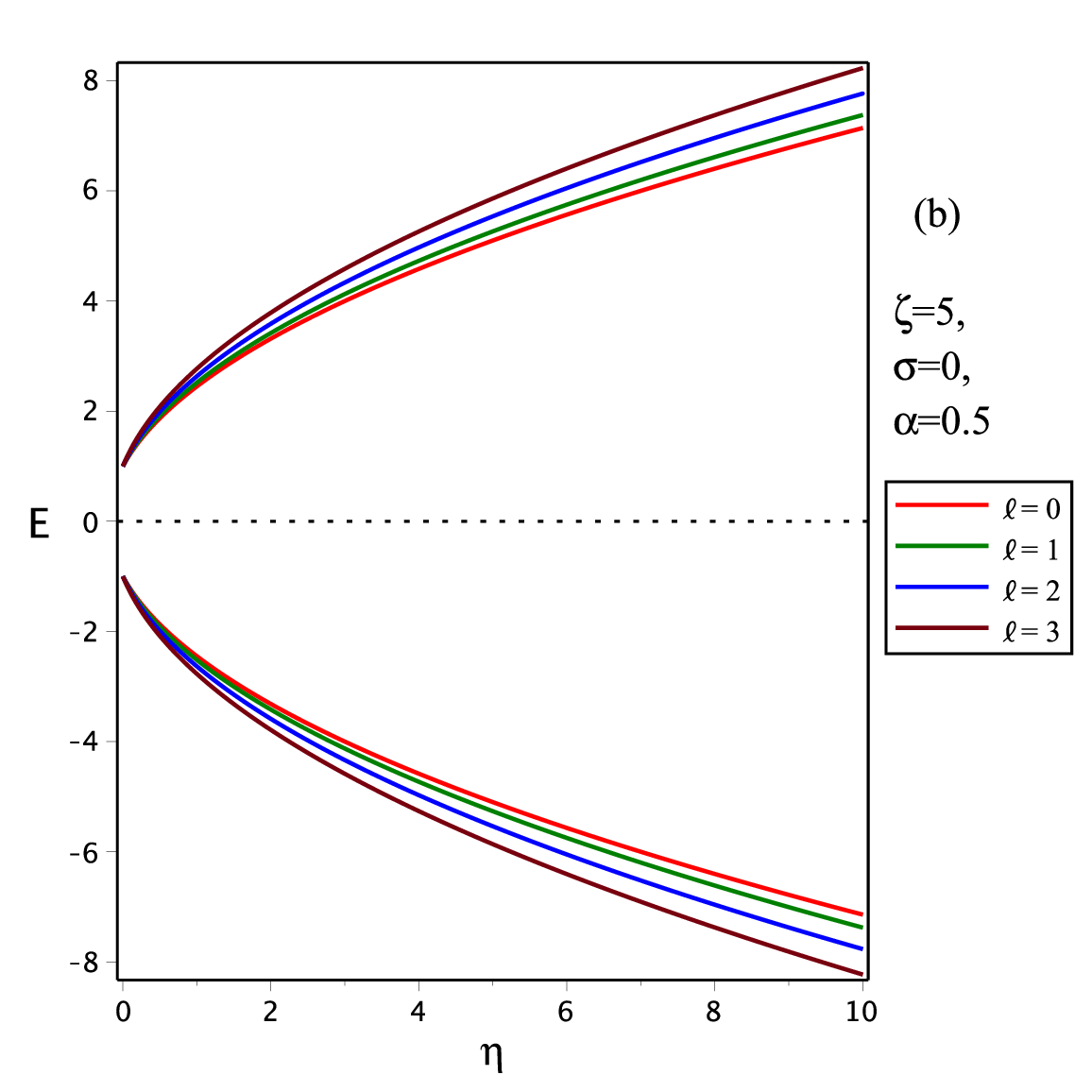} 
\includegraphics[width=0.3\textwidth]{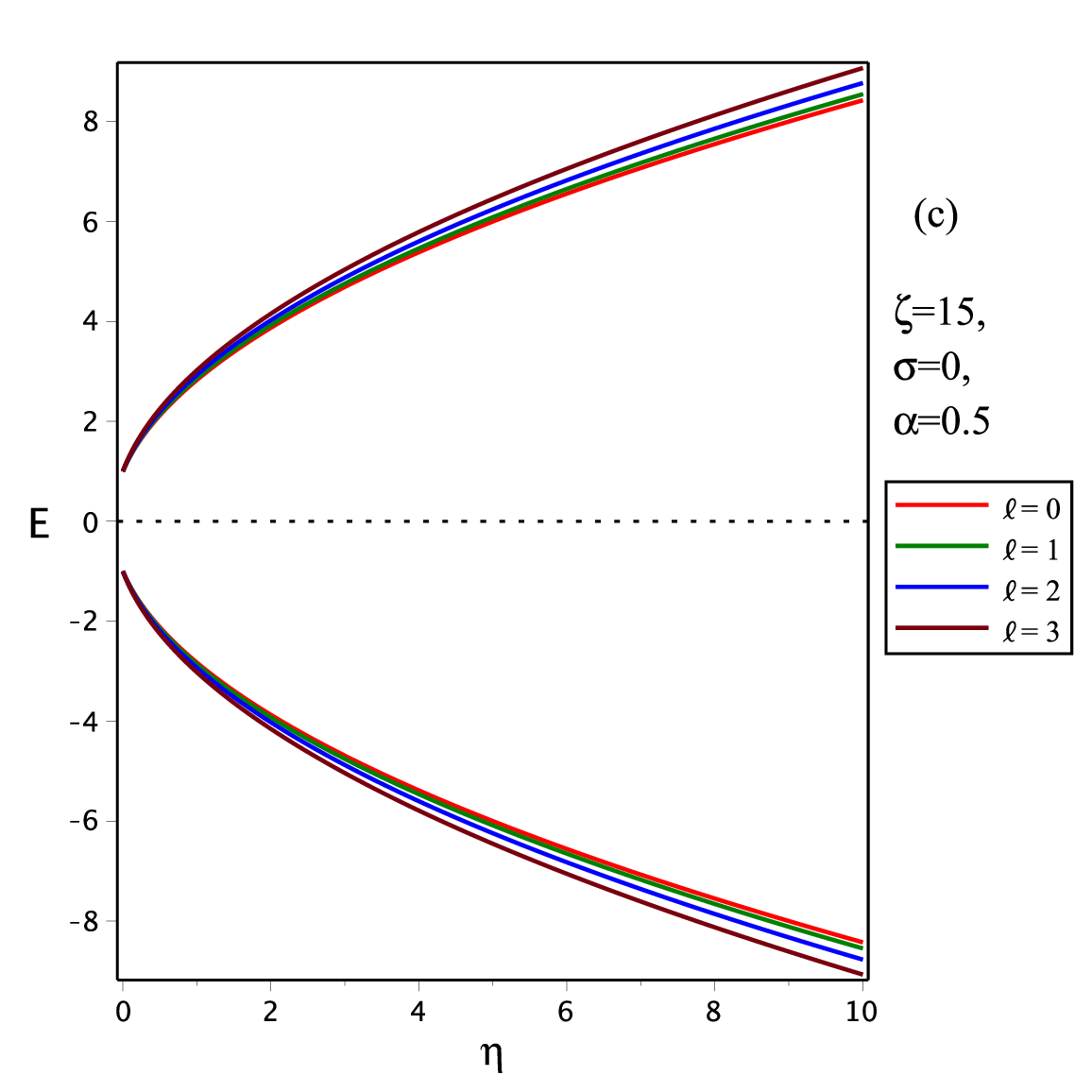}
\caption{\small 
{The energy levels Eq.(\ref{93}) of the KG-oscillators in a
PGM/QPGM spacetime and a Wu-Yang magnetic monopole for $n_{r}=1$, $\ell
=0,1,2,3$, (a) for $\zeta =0.5$, (b) for $\zeta =5$, and (c) for $\zeta =15$.}}
\label{fig2}
\end{figure}%

In Figures 1, 2, and 3, we plot the energy levels, Eq.(\ref{93}), against the oscillator frequency $\eta $. Figures 1(a),1(b), and 1(c), are plotted for $\alpha =1$, $\alpha =0.5$, and $\alpha =0.05$, respectively, to observe the effect of the PGM/QPGM parameter $\alpha $ (where we take $\zeta =0$, the coupling constant of the scalar curvature $R_{\mu }^{\mu }$ , and $%
\sigma =0$ for the WYMM strength) on the energy levels of KG-oscillators in a PGM?QPGM spacetime and a WYMM at $%
S\left( r\right) =0$, $\mathcal{F}_{r}\left( r\right) =\eta \,r$. It can be clearly observed that as $\alpha $ decreases from $\alpha =1$ towards $\alpha \sim 0$, the energy levels are destined to cluster on $E=\pm m_{\circ }$, as is obvious from the asymptotic tendency of (\ref{93}) at $\alpha \sim 0$. Moreover, as the PGM parameter $\alpha $ increases from above zero value towards $\alpha =1$ (i.e., flat Minkowski spacetime) the particles/anti-particles' energies increase with an increasing frequency $%
\eta $. In Figures 2(a), 2(b), and 2(c) we plot the energy levels Eq.(\ref%
{93}) against the oscillator frequency $\eta $ at a fixed $\alpha =0.5$ and $%
\sigma =0$ for $\zeta =0.5$, $\zeta =5$, and $\zeta =15$, respectively. Hereby, we observe that not only the energies for the particles and anti-particles increase with increasing coupling constant $\zeta $ of the scalar curvature $R_{\mu }^{\mu }$, but we also observe eminent clustering of the energy levels.
\begin{figure}[!ht]  
\centering
\includegraphics[width=0.3\textwidth]{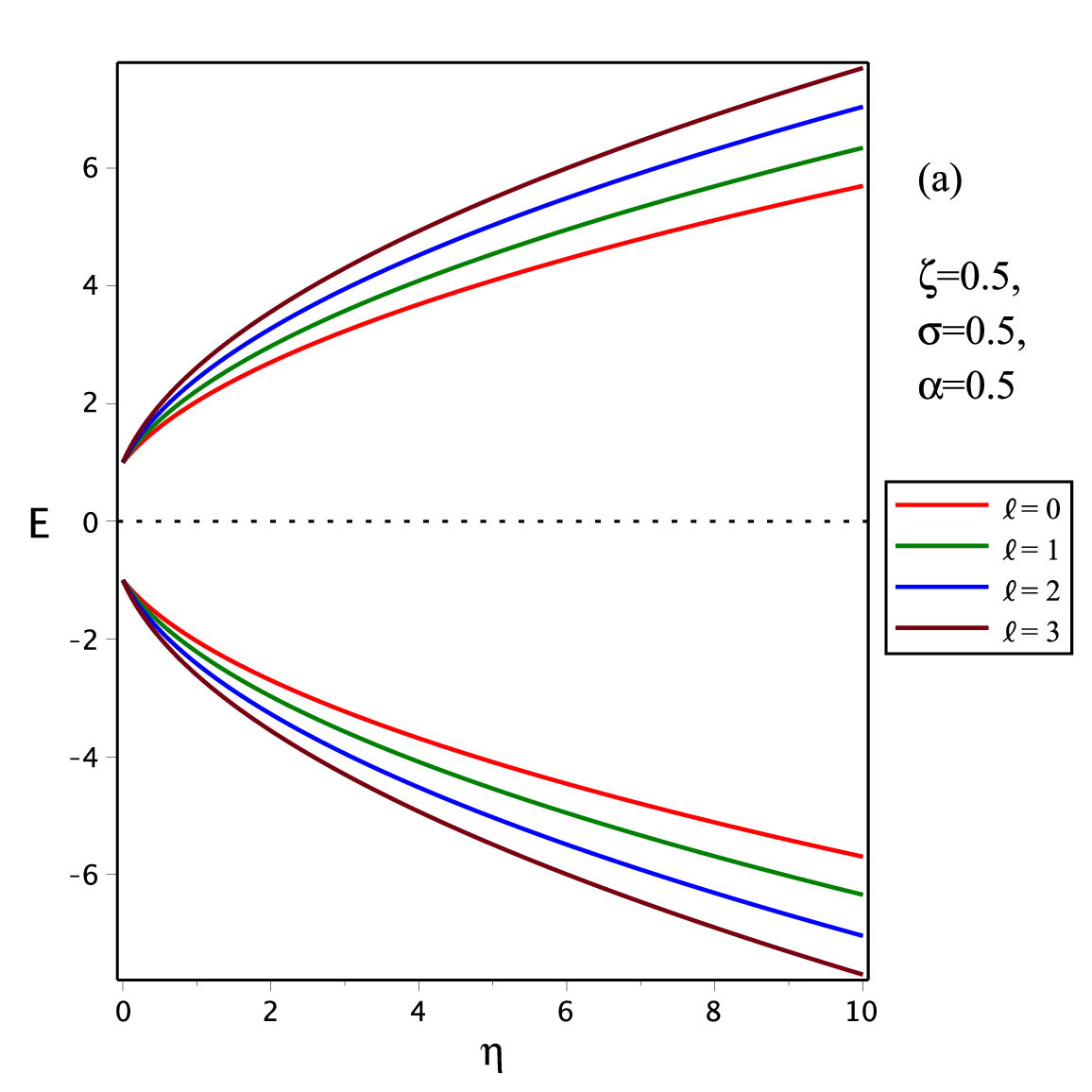}
\includegraphics[width=0.3\textwidth]{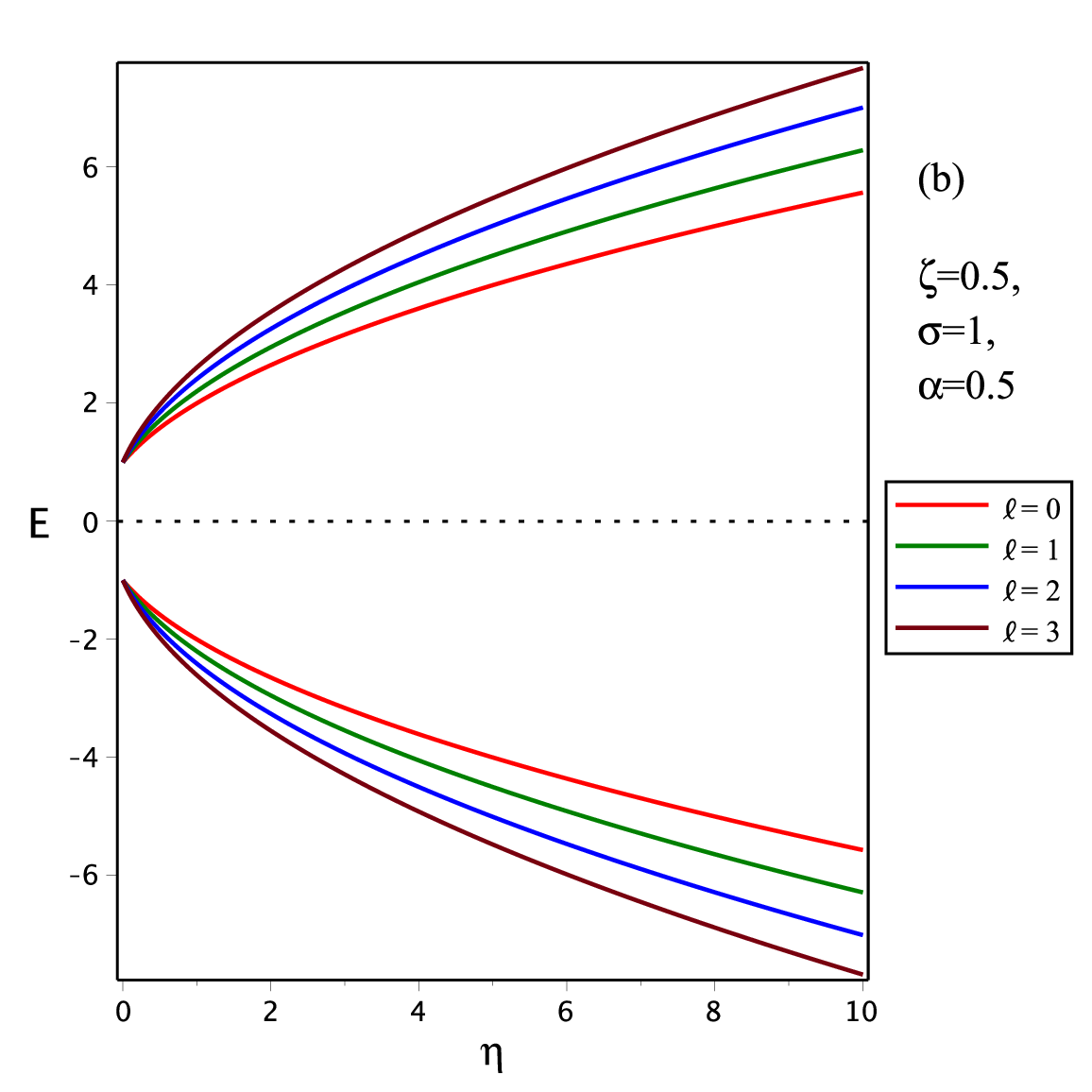} 
\includegraphics[width=0.3\textwidth]{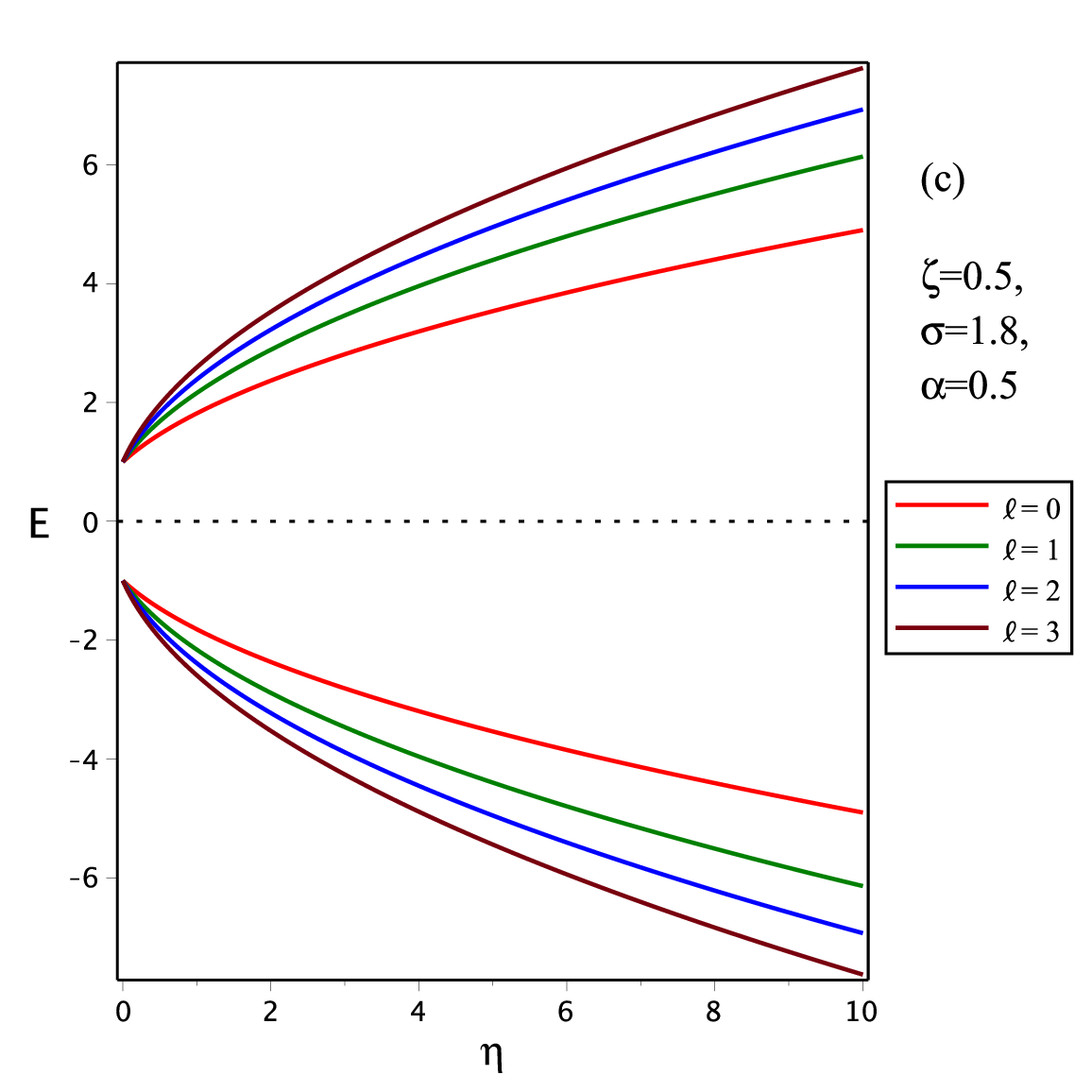}
\caption{\small 
{ The energy levels Eq.(\ref{93}) of the KG-oscillators in a
PGM/QPGM spacetime and a Wu-Yang magnetic monopole for $n_{r}=1$, $\ell
=0,1,2,3$, (a) for $\sigma =0.5$, (b) for $\sigma =1$, and (c) for $\sigma=1.8$.}}
\label{fig3}
\end{figure}%
In Figures 3(a), 3(b), and 3(c), we plot the energy levels Eq.(\ref{93}) against the oscillator frequency $\eta $ at the fixed values $\alpha =0.5$ and $\zeta =0.5$ for $\sigma =0.5$, $\sigma =1$, and $%
\sigma =1.8$, respectively. It is clear that the separation between the energy levels increases with increasing $\sigma $.

At this point, nevertheless, one should be reminded that the KG-oscillators in Eq.(\ref{9}) (i.e., KG-oscillators in a PGM spacetime Eq.(\ref{e01}) and a WYMM) are isospectral and share the same energy levels, Eq.(\ref{93}), with those in Eq.(\ref{e0795}) (i.e., KG-oscillators in a QPGM spacetime in Eq.(\ref{e03}) and a WYMM). Two
illustrative examples are in order.

\subsection{Exponential type metric function $q\left(\rho \right)
=e^{a\rho ^{2}}$}

Let us take a positive-valued (non-zero) deformation/transformation function $q\left( \rho \right) =e^{a\rho ^{2}}$ to yield, through (\ref{e02}), $%
f\left( \rho \right) =$ $\left( 1+a\rho ^{2}\right) e^{a\rho ^{2}}$. This would, with $\mathcal{F}_{r}\left( r\right) =\eta \,r=\eta \sqrt{q\left(
\rho \right) }\rho =$ $\mathcal{F}_{\rho }\left( \rho \right) /\sqrt{f\left(
\rho \right) }$, consequently imply that $\mathcal{F}_{\rho }\left( \rho
\right) =\eta \left( \rho +a\rho ^{3}\right) e^{a\rho ^{2}}$. Our QPGM spacetime (\ref{e03}) now reads%
\begin{equation}
ds^{2}=-dt^{2}+\frac{\left( 1+a\rho ^{2}\right) e^{a\rho ^{2}}}{\alpha ^{2}}%
d\rho ^{2}+e^{a\rho ^{2}}\,\rho ^{2}\left[ d\theta ^{2}+\sin ^{2}\theta
\,d\varphi ^{2}\right] ,  \label{931}
\end{equation}%
which consequently implies that the KG-particles in equation (\ref{e0795}) are now described by%
\begin{gather}
\left\{ \frac{\alpha ^{2}}{\sqrt{\left( 1+a\rho ^{2}\right) }e^{3a\rho
^{2}/2}\rho ^{2}}\,\left( \partial _{\rho }+\eta \left( \rho +a\rho
^{3}\right) e^{a\rho ^{2}}\right) \,\frac{e^{a\rho ^{2}/2}\rho ^{2}}{\sqrt{%
\left( 1+a\rho ^{2}\right) }}\,\,\left( \partial _{\rho }-\eta \left( \rho
+a\rho ^{3}\right) e^{a\rho ^{2}}\right) \right.   \notag \\
\left. -\frac{L\left( L+1\right) }{e^{a\rho ^{2}}\,\rho ^{2}}+\left[
E^{2}-m_{\circ }^{2}\right] \right\} \psi \left( \rho \right) =0.
\label{932}
\end{gather}%
This is nothings but the KG-oscillators in a PGM spacetime and a WYMM as described by (\ref{9}) and admits the exact solvability in (\ref{91}) and (\ref{93}). That is, the two KG-systems in Eq. (\ref{9}) and (\ref%
{932}) are invariant and isospectral.

\subsection{Power-law type metric function $q\left(\rho \right) =b%
\rho ^{\upsilon }$}

Following the same procedure as above, we obtain $f\left( \rho \right)
=\left( 1+\upsilon /2\right) ^{2}b\rho ^{\upsilon },$ $\upsilon \neq 0,-2$
and $\mathcal{F}_{\rho }\left( \rho \right) =\eta b\left( 1+\upsilon
/2\right) \rho ^{\upsilon +1}$. At this point, one should notice that the restrictions on $\upsilon $ are unavoidable to secure positive-valued characterization for the deformation/transformation functions (metric functions) $q\left( \rho
\right) $ and $f\left( \rho \right) $. Our QPGM spacetime of (\ref{e03}) now reads%
\begin{equation}
ds^{2}=-dt^{2}+\frac{\left( 1+\upsilon /2\right) ^{2}b\rho ^{\upsilon }}{%
\alpha ^{2}}d\rho ^{2}+b\rho ^{\upsilon +2}\left[ d\theta ^{2}+\sin
^{2}\theta \,d\varphi ^{2}\right] ,  \label{933}
\end{equation}%
and consequently yields that the KG-particles in equation (\ref{e0795}) are now described by%
\begin{gather}
\left\{ \frac{\alpha ^{2}}{\left( 1+\upsilon /2\right) b^{3/2}\rho
^{3\upsilon /2+2}}\,\left( \partial _{\rho }+\eta b\left( 1+\upsilon
/2\right) \rho ^{\upsilon +1}\right) \,\frac{b^{1/2}\rho ^{\upsilon /2+2}}{%
\left( 1+\upsilon /2\right) }\,\,\left( \partial _{\rho }-\eta b\left(
1+\upsilon /2\right) \rho ^{\upsilon +1}\right) \right.   \notag \\
\left. -\frac{L\left( L+1\right) }{b\rho ^{\upsilon +2}}+\left[
E^{2}-m_{\circ }^{2}\right] \right\} \psi \left( \rho \right) =0.
\label{934}
\end{gather}

For both illustrative examples above, one should notice that the two KG-oscillators in  (\ref{932}) and (\ref{934}) in the QPGM spacetime of (\ref%
{e03}) are invariant and isospectral with the KG-oscillators in (\ref{9}) in the PGM spacetime of (\ref{e01}). 

\section{KG-Coulombic particles in a QPGM spacetime and a WYMM with a Bessel-type $\mathcal{F}_{r}\left( r\right) $}

In this section, we consider KG-Coulomb like particles a QPGM spacetime and a WYMM at $S\left( r\right) =0$. The substitution of a Bessel functions type%
\begin{equation}
\mathcal{F}_{r}=\mathcal{F}_{r}\left( r\right) =-\frac{A\,J_{2}\left( \sqrt{%
4Ar}\right) }{\sqrt{Ar}J_{1}\left( \sqrt{4Ar}\right) },  \label{935}
\end{equation}%
in Eq.(\ref{e08}) would result%
\begin{equation}
\left\{ \partial _{r}^{2}-\frac{\tilde{L}\left( \tilde{L}+1\right) }{r^{2}}+%
\frac{A}{r}+\Lambda \right\} R\left( r\right) =0.  \label{936}
\end{equation}%
This is a Schr\"{o}dinger-like radial equation that admits an exact textbook solution%
\begin{equation}
R\left( r\right) =C\,e^{-i\sqrt{\Lambda }r}\,r^{\tilde{L}+1}\,\,U\left( 
\tilde{L}+1+\frac{iA}{\sqrt{\Lambda }},2\tilde{L}+2,2i\sqrt{\Lambda }%
r\right) .  \label{937}
\end{equation}%
We again use the condition that the confluent hypergeometric series $U\left(
a,b,z\right) $ is truncated into a polynomial of order $n_{r}=0,1,2,\cdots $ so that%
\begin{equation}
\tilde{L}+1+\frac{iA}{\sqrt{\Lambda }}=-n_{r}\Longleftrightarrow \Lambda =-%
\frac{A^{2}}{\left( n_{r}+\tilde{L}+1\right) ^{2}}\Longrightarrow E=\pm 
\sqrt{m_{\circ }^{2}-\frac{\alpha ^{2}A^{2}}{\left( n_{r}+\tilde{L}+1\right)
^{2}}}.  \label{938}
\end{equation}
At this point, one should observe that The KG-Coulombic particles in (\ref%
{936}) are invariant and isospectral with those in (\ref{e0795}). That is, they are invariant and isospectral with the KG-particles in   
\begin{equation}
\left\{ \frac{1}{\sqrt{f\left( \rho \right) }q\left( \rho \right) \rho ^{2}}%
\,\left( \partial _{\rho }+\mathcal{F}_{\rho }\left( \rho \right) \right) \,%
\frac{q\left( \rho \right) \rho ^{2}}{\sqrt{f\left( \rho \right) }}%
\,\,\left( \partial _{\rho }-\mathcal{F}_{\rho }\left( \rho \right) \right) -%
\frac{\tilde{L}\left( \tilde{L}+1\right) }{q\left( \rho \right) \,\rho ^{2}}%
+\Lambda \right\} \psi \left( \rho \right) =0;\;\Lambda =\frac{%
E^{2}-m_{\circ }^{2}}{\alpha ^{2}}.  \label{939}
\end{equation}%
Hereby, $\mathcal{F}_{\rho }\left( \rho \right) =\mathcal{F}_{r}\left(
r\left( \rho \right) \right) \sqrt{f\left( \rho \right) }$ would read%
\begin{equation}
\mathcal{F}_{\rho }\left( \rho \right) =-\frac{A\,J_{2}\left( \sqrt{%
4Ar\left( \rho \right) }\right) }{\sqrt{Ar\left( \rho \right) }J_{1}\left( 
\sqrt{4Ar\left( \rho \right) }\right) }\sqrt{f\left( \rho \right) },
\label{9391}
\end{equation}%
\begin{figure}[!ht]  
\centering
\includegraphics[width=0.3\textwidth]{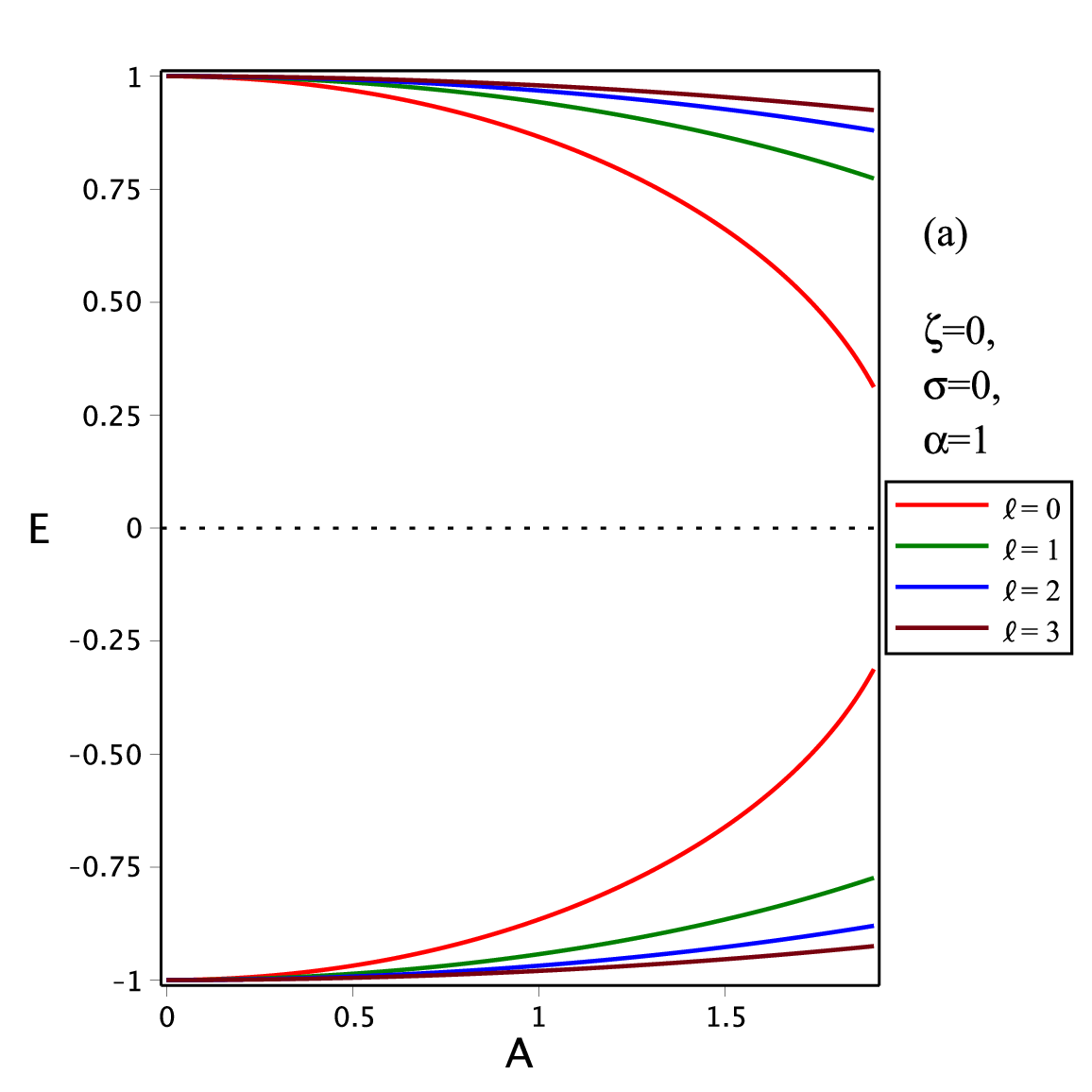}
\includegraphics[width=0.3\textwidth]{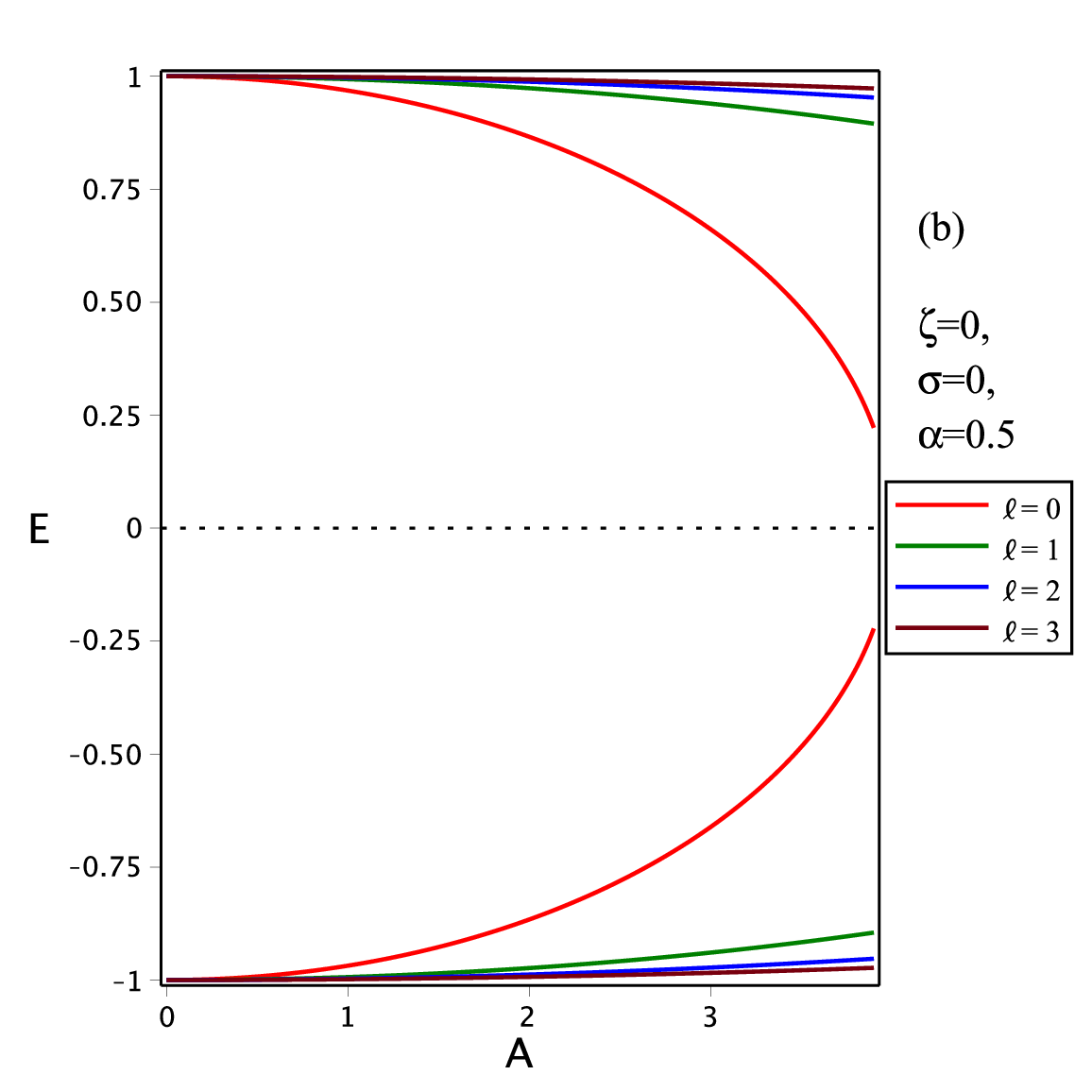} 
\includegraphics[width=0.3\textwidth]{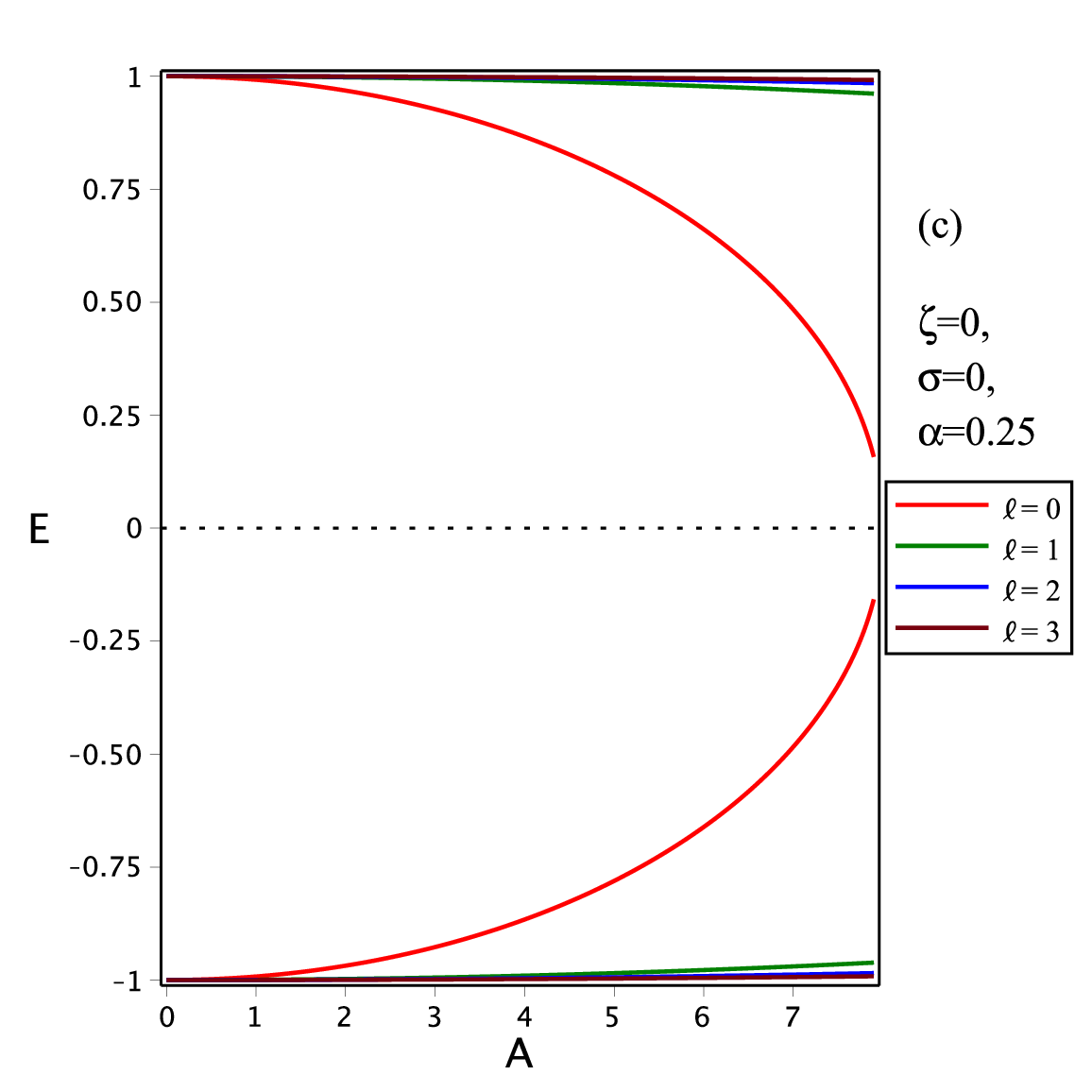}
\caption{\small 
{ The energy levels Eq.(\ref{938}) of the KG-Coulombic
particles in a PGM/QPGM spacetime and a Wu-Yang magnetic monopole for $%
n_{r}=1$, $\ell =0,1,2,3$, (a) for $\alpha =1$, (b) for $\alpha =0.5$, and (c) for $\alpha =0.25$.}}
\label{fig4}
\end{figure}%
\begin{figure}[!ht]  
\centering
\includegraphics[width=0.3\textwidth]{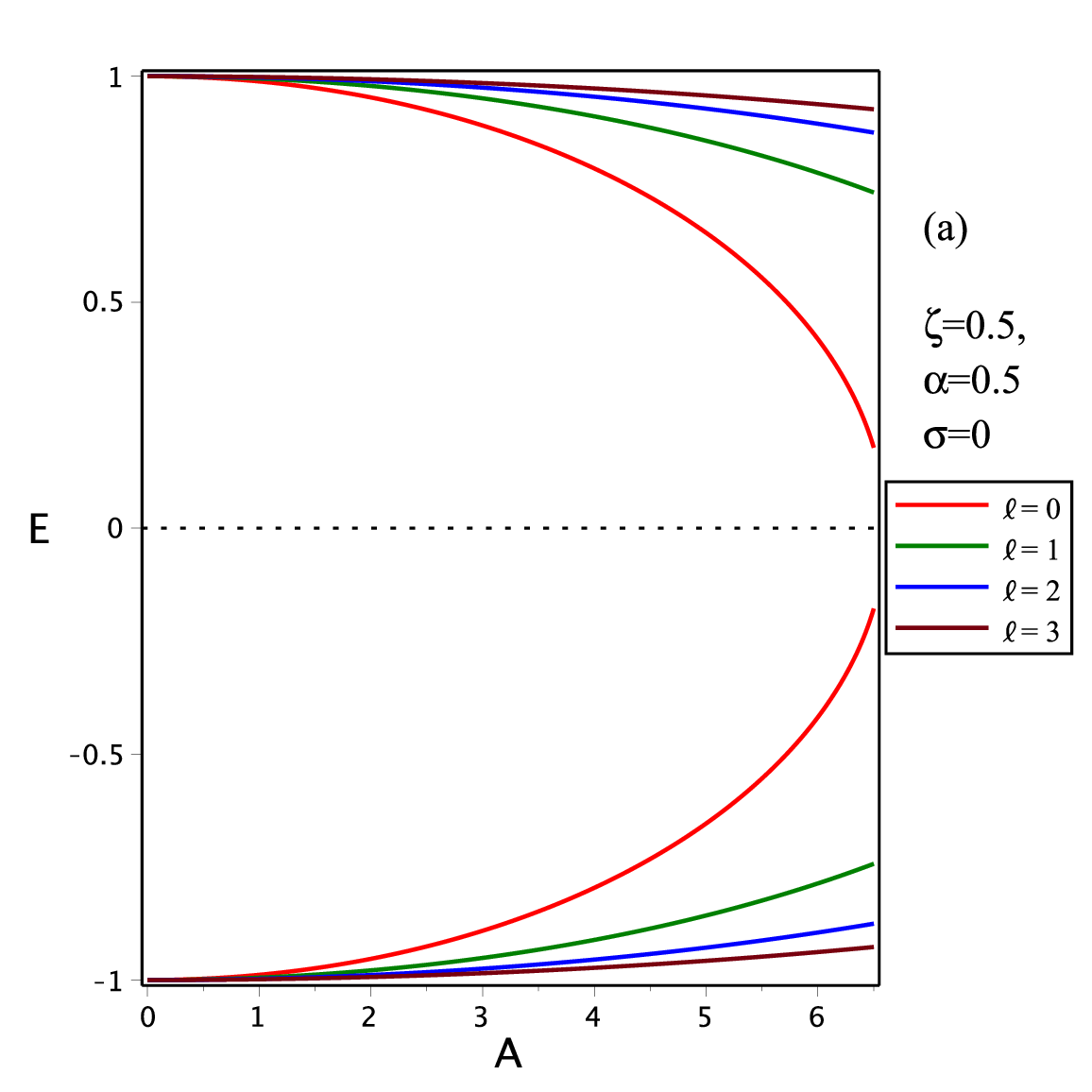}
\includegraphics[width=0.3\textwidth]{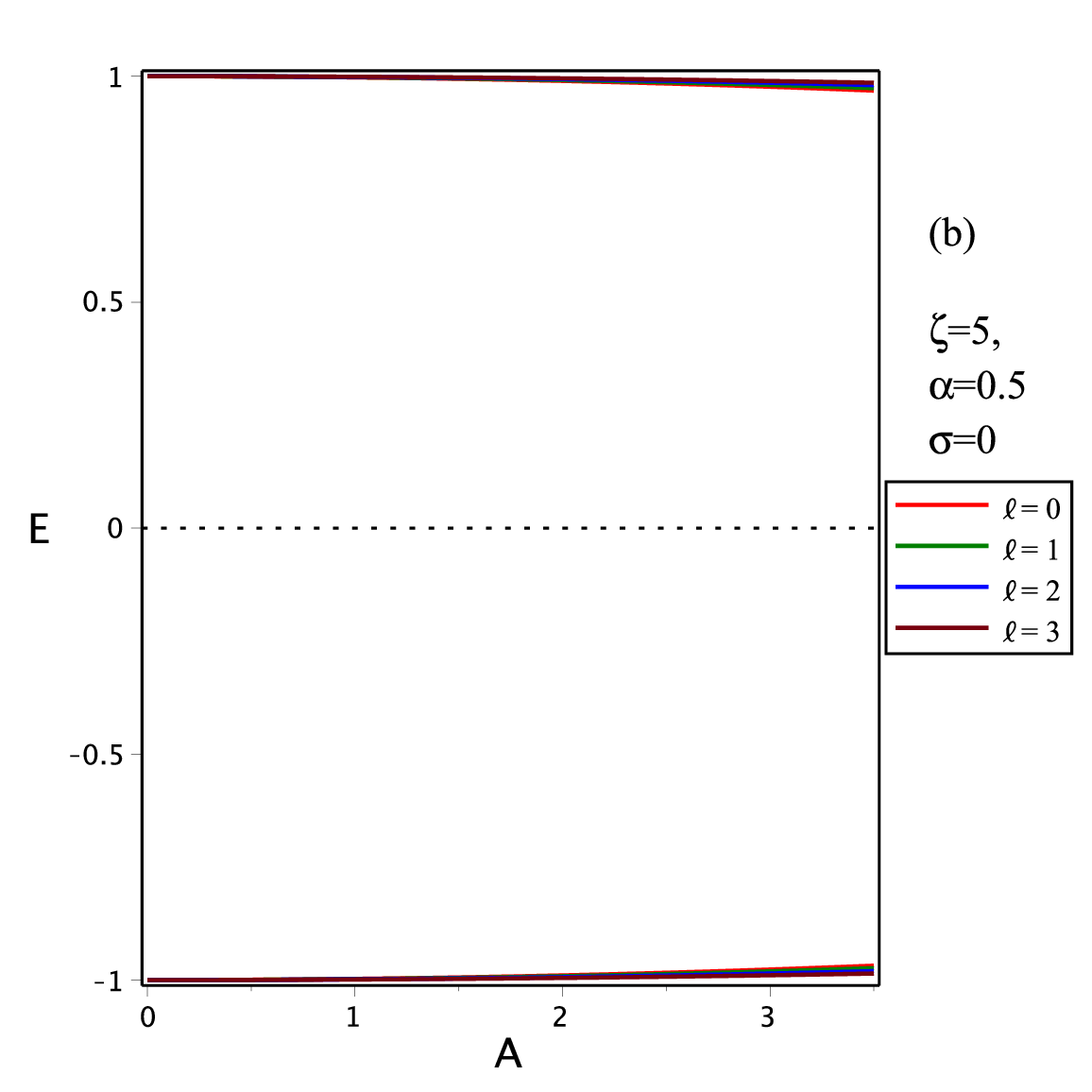} 
\includegraphics[width=0.3\textwidth]{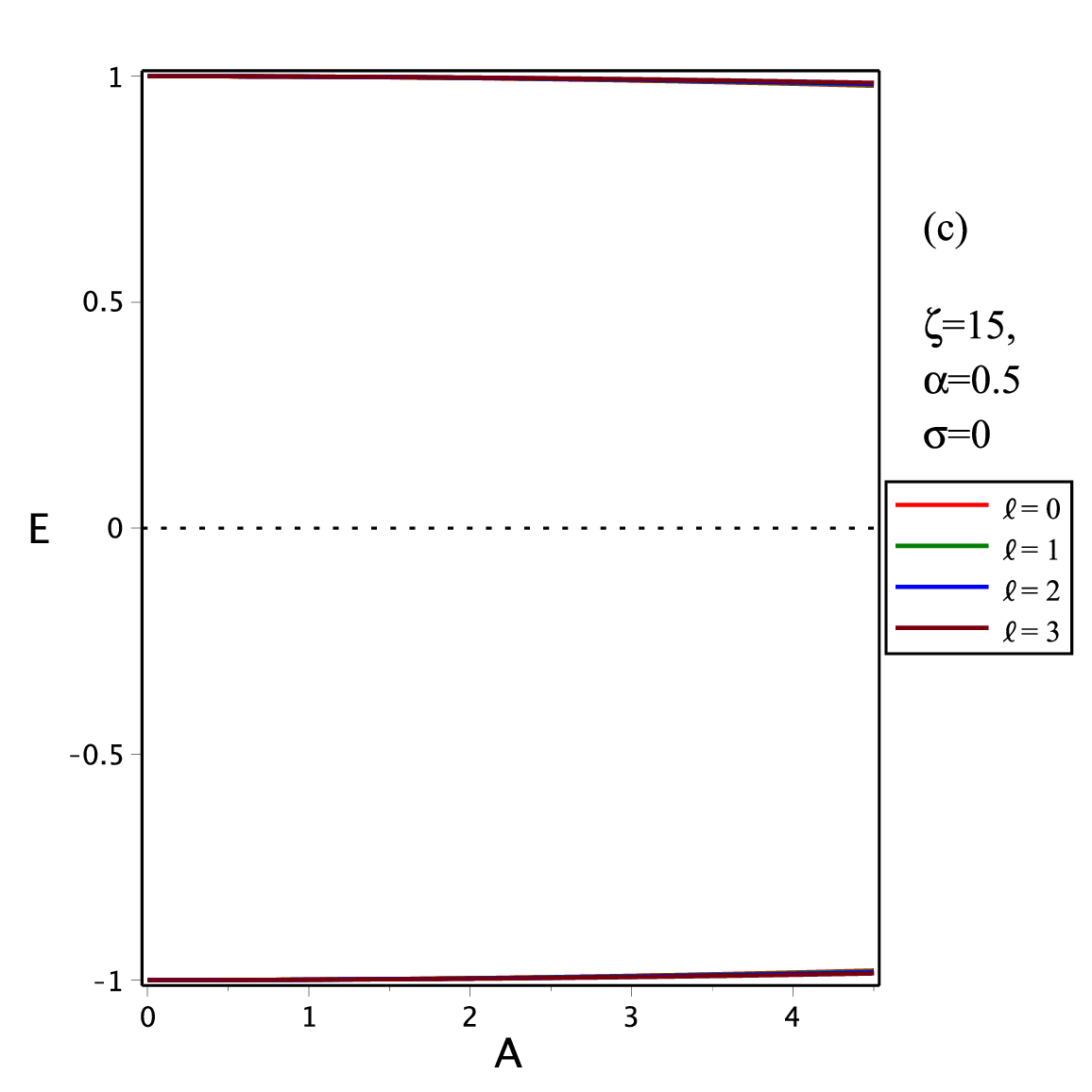}
\caption{\small 
{ The energy levels Eq.(\ref{938}) of the KG-Coulombic particles
in a PGM/QPGM spacetime and a Wu-Yang magnetic monopole for $n_{r}=1$, $\ell=0,1,2,3$, (a) for $\zeta =0.5$, (b) for $\zeta =5$, and (c) for $\zeta =15$}}
\label{fig5}
\end{figure}%
\begin{figure}[!ht]  
\centering
\includegraphics[width=0.3\textwidth]{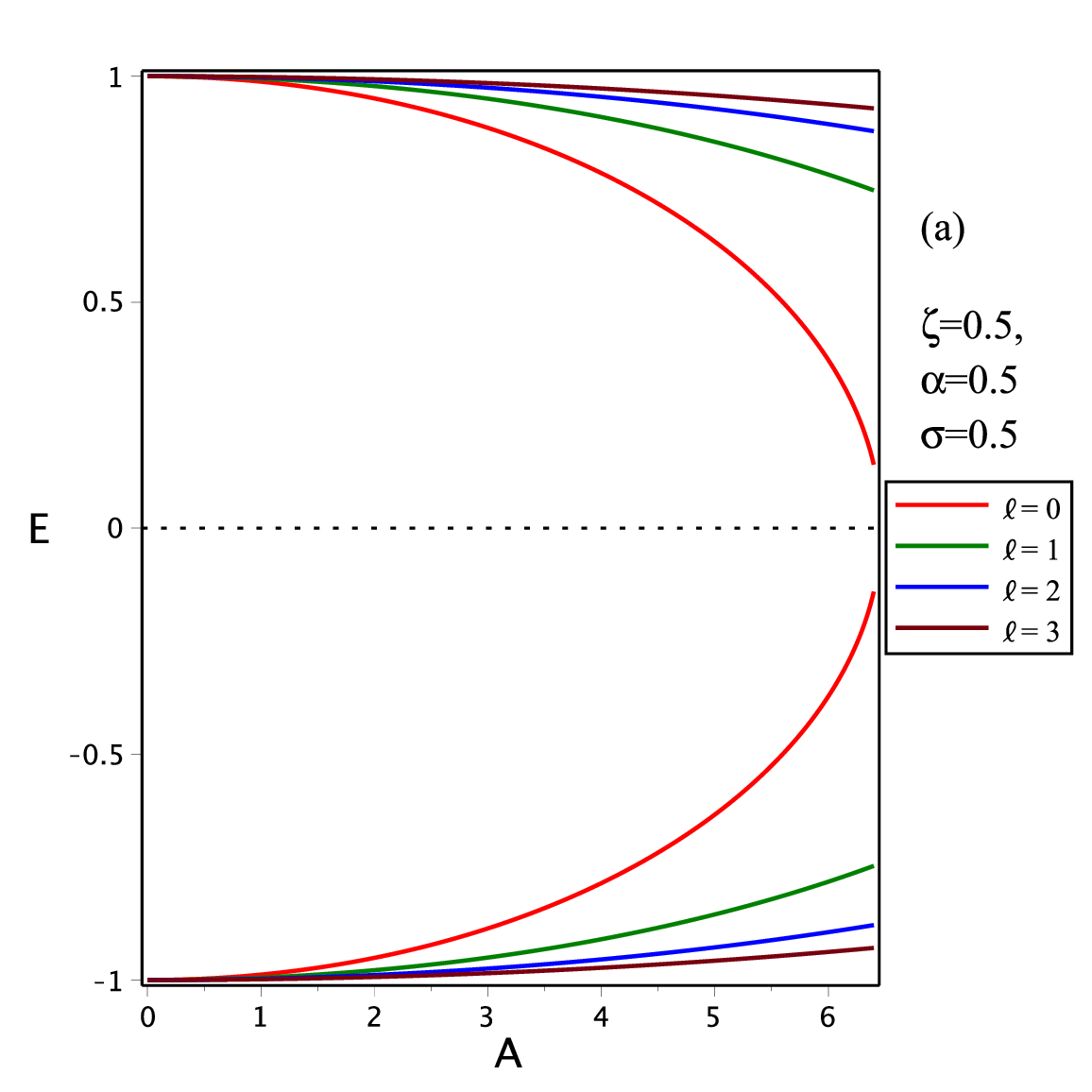}
\includegraphics[width=0.3\textwidth]{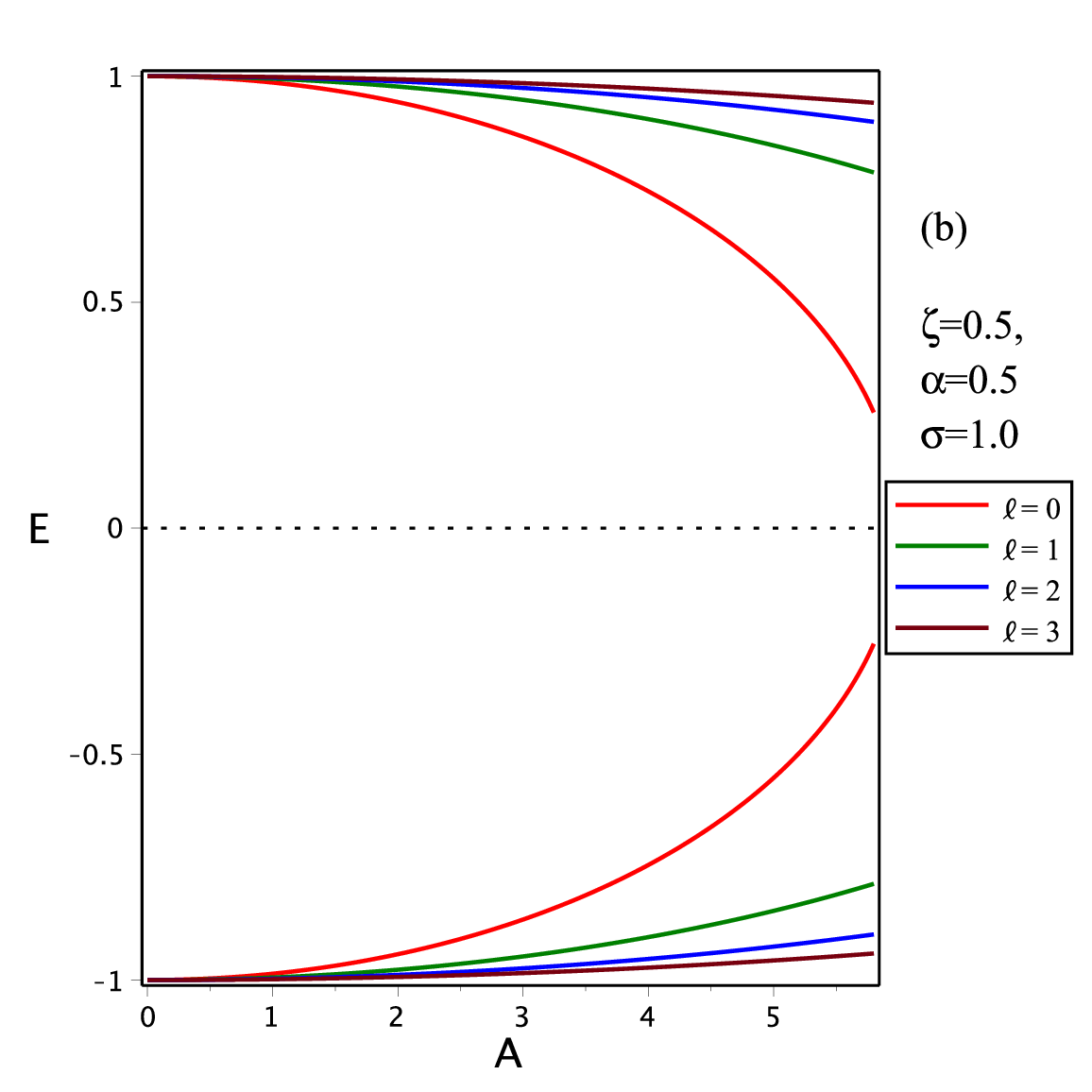} 
\includegraphics[width=0.3\textwidth]{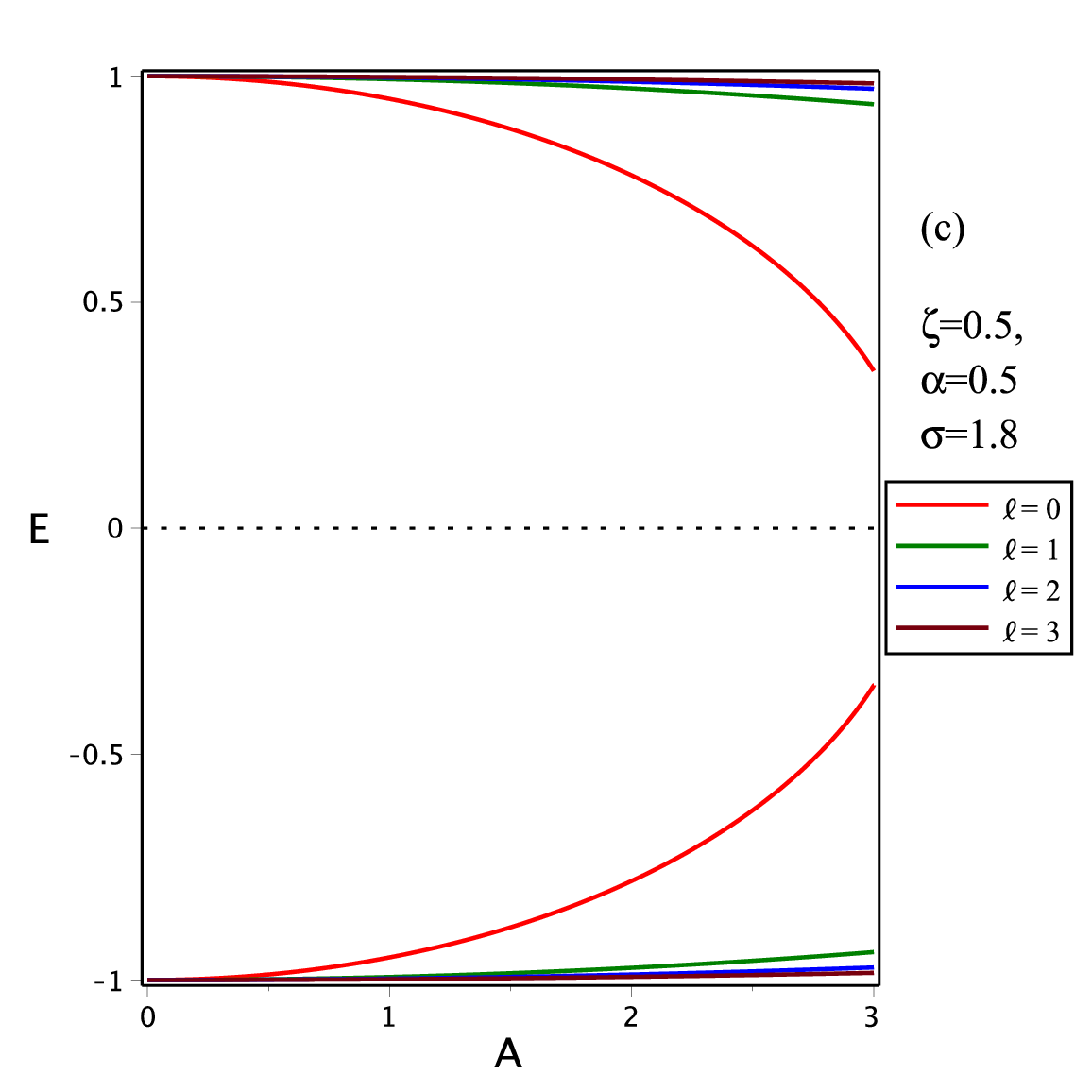}
\caption{\small 
{ The energy levels Eq.(\ref{938}) of the KG-Coulombic particles
in a PGM/QPGM spacetime and a Wu-Yang magnetic monopole for $n_{r}=1$, $\ell
=0,1,2,3$, (a) for $\sigma =0.5$, (b) for $\sigma =1$, and (c) for $\sigma=1.8$.}}
\label{fig6}
\end{figure}%
where, for the deformation/transformation functions discussed in the preceding section,%
\begin{equation}
r\left( \rho \right) =\left\{ 
\begin{tabular}{ll}
$\rho \,e^{a\rho ^{2}/2}$ & for $q\left( \rho \right) =e^{a\rho
^{2}}\Longleftrightarrow $ $f\left( \rho \right) =$ $\left( 1+a\rho
^{2}\right) e^{a\rho ^{2}}\medskip $ \\ 
$\sqrt{b}\rho ^{1+\upsilon /2}$ & for $q\left( \rho \right) =b\rho
^{\upsilon }\Longleftrightarrow f\left( \rho \right) =\left( 1+\upsilon
/2\right) ^{2}b\rho ^{\upsilon }\medskip $%
\end{tabular}%
\right. .  \label{9392}
\end{equation}%
One should also notice that equation (\ref{939}), along with (\ref{9391}), describes KG-Coulombic particles in the QPGM spacetime of (\ref{e03}) and equation (\ref{936}) describes KG-Coulombic particles in the regular PGM spacetime (\ref{e01}).

In Figures 4, 5, and 6, we plot the energy levels Eq. (\ref{938}) of the KG-Coulombic particles in a PGM/QPGM spacetime and a WYMM. In Figures 4(a),4(b), and 4(c), are plotted for $\alpha =1$, $\alpha =0.5$, and $\alpha =0.25$, respectively, to observe the effect of global monopole parameter $\alpha $. We may, again, observe that as the PGM
parameter decreases from $\alpha =1$ (i.e., Minkowski flat spacetime) the energy levels tend to cluster around $E=\pm m_{\circ }$. This is obvious from equation (\ref{938}) where a Taylor series expansion about $\alpha =0$
would yield $E_{\alpha \rightarrow 0}=\pm \left[ m_{\circ }+O\left( \alpha
^{4}\right) \right] .$  Moreover, the separation between energy levels decreases (with a tendency of clustering) with increasing WYMM parameter $\sigma$.

\section{KG-Coulomb particles in a PGM spacetime and a WYMM at $S( r) =K/r$, $\mathcal{F}_{r} (r) =0.$}

It has been shown by de Mello and Furtado \cite{Re133} that a self-interaction \ on a charged particle placed in a PGM spacetime is given by $S\left( r\right) =K\left( \alpha \right) /r$, where $K\left( \alpha \right) $ is a positive constant and $r$ is the distance to the PGM. For the sake of simplicity of notations, let us consider that the Lorentz scalar potential $S\left( r\right) =K/r$ and $\mathcal{F}%
_{r}\left( r\right) =0$ in (\ref{e08}) to obtain, in a straightforward
manner, that 
\begin{equation}
\left\{ \partial _{r}^{2}-\frac{\mathcal{L}\left( \mathcal{L}+1\right) }{%
r^{2}}-\frac{2B}{r}+\Lambda \right\} R\left( r\right) =0,  \label{101}
\end{equation}%
where, $B=\tilde{m}\mathcal{\tilde{K}},$ $\mathcal{\tilde{K}}=K/\alpha $ and$\;$%
\begin{equation}
\mathcal{L}\left( \mathcal{L}+1\right) =\tilde{L}\left( \tilde{L}+1\right) +%
\mathcal{\tilde{K}}^{2}\Longleftrightarrow \mathcal{L=-}\frac{1}{2}+\sqrt{%
\frac{1}{4}+\tilde{L}\left( \tilde{L}+1\right) +\mathcal{\tilde{K}}^{2}}.
\label{101.1}
\end{equation}%
This system admits an exact textbook solution in the form of confluent hypergeometric series and reads%
\begin{equation}
R\left( r\right) =C\,e^{-i\sqrt{\Lambda }r}\left( 2i\sqrt{\Lambda }r\right)
^{\mathcal{L}+1}\,\,U\left( \mathcal{L}+1-\frac{iB}{\sqrt{\Lambda }},2%
\mathcal{L}+2,2i\sqrt{\Lambda }r\right) .  \label{102}
\end{equation}%
Again, finiteness and square integrability of the radial function $R\left(
r\right) $ enforces the truncation of the confluent hypergeometric series of the second kind $U\left(a,b,z\right) $ into a polynomial of order $n_{r}$ so that%
\begin{equation}
\mathcal{L}+1-\frac{iB}{\sqrt{\Lambda }}=-n_{r}\Longleftrightarrow i\sqrt{%
\Lambda }=-\frac{B}{\tilde{n}}\,\Longleftrightarrow \Lambda =-\frac{B^{2}}{%
\tilde{n}^{2}},  \label{103}
\end{equation}%
where $\tilde{n}=n_{r}+\mathcal{L}+1$ is an irrational principle quantum number (which collapses into the usual principle quantum number for $\alpha =1
$ and $\sigma =0$). As a result, we get the energies and the radial function $R\left( r\right) $ of such KG-Coulombic particles in a PGM spacetime and WYMM as%
\begin{equation}
E=\pm \sqrt{m_{\circ }^{2}-\frac{\alpha ^{2}B^{2}}{\tilde{n}^{2}}}=\pm \sqrt{%
m_{\circ }^{2}-\frac{m_{\circ }^{2}\,K^{2}}{\alpha ^{2}\left(
n_{r}+\mathcal{L}+1\right) ^{2}}},  \label{104}
\end{equation}%
and%
\begin{equation}
R\left( r\right) =\tilde{C}\exp (-\frac{Br}{\tilde{n}})\;r^{\mathcal{L}%
+1}\,\,U\left( -n_{r},2\mathcal{L}+2,2\frac{B}{\tilde{n}}r\right) ;\;B=\frac{%
m_{\circ }K}{\alpha ^{2}},  \label{105}
\end{equation}%
\begin{figure}[!ht]  
\centering
\includegraphics[width=0.3\textwidth]{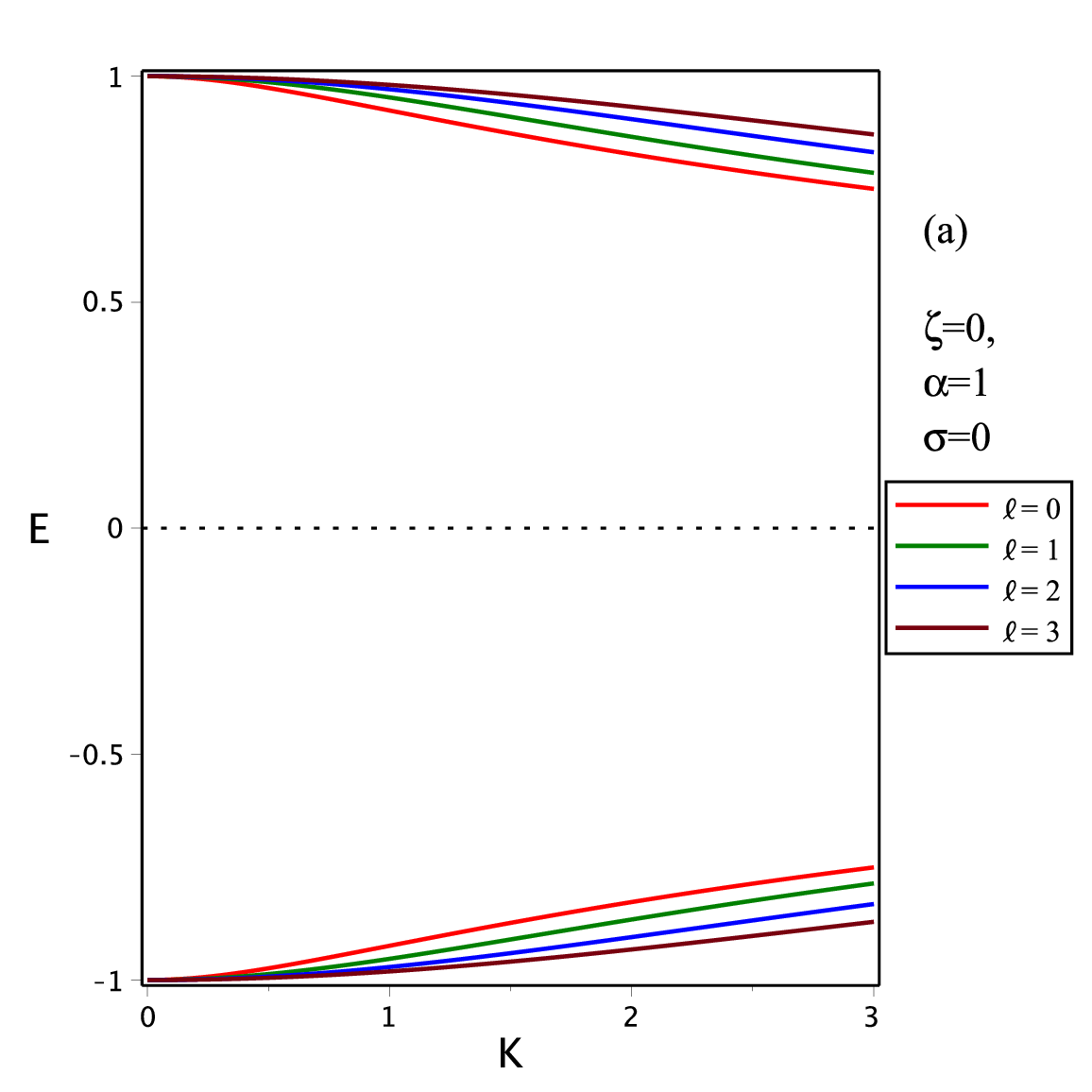}
\includegraphics[width=0.3\textwidth]{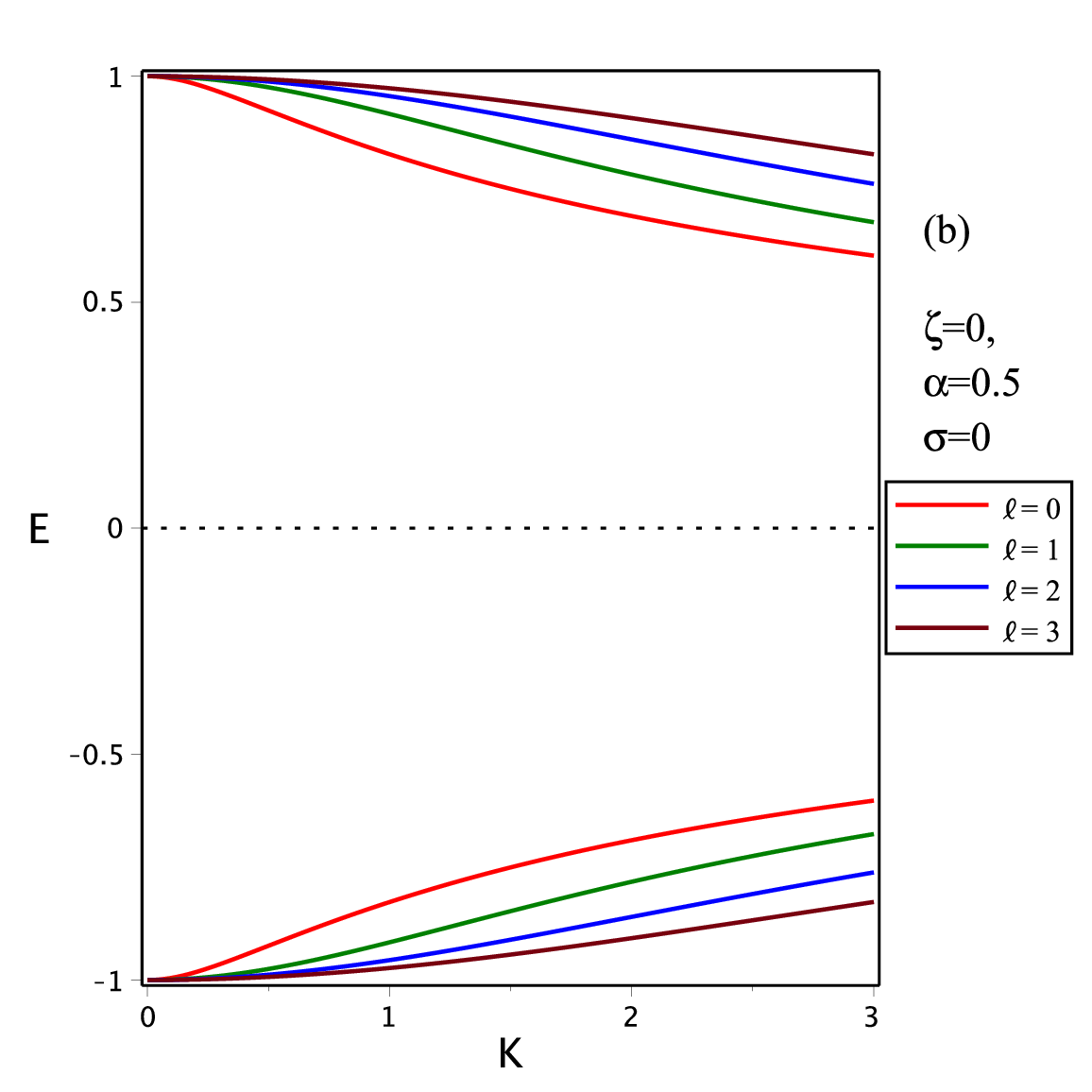} 
\includegraphics[width=0.3\textwidth]{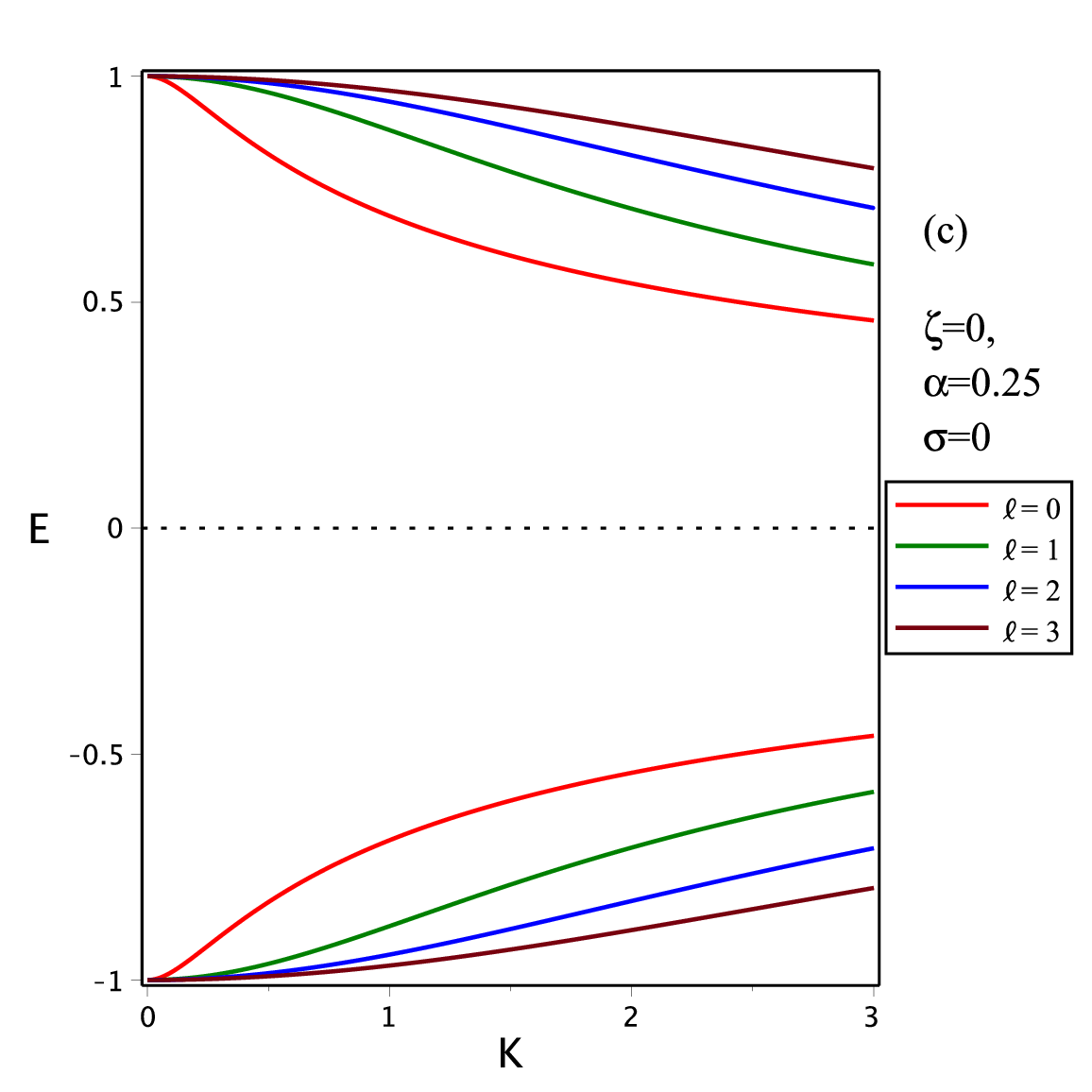}
\caption{\small 
{ The energy levels Eq. (\ref{105}) of the  KG-Coulombic
particles in a PGM spacetime and a Wu-Yang magnetic monopole for $n_{r}=1$, $%
\ell =0,1,2,3$, (a) for $\alpha =1$, (b) for $\alpha =0.5$, and (c) for $\alpha =0.25$.}}
\label{fig7}
\end{figure}%
respectively, where $\mathcal{L}$ is given by Eq.(\ref{101.1}). At this point, one should observe that this result is in exact accord with that reported by de Oliveira and de Mello \cite{Re8} (i.e., their result in Eq. (76) with their $\alpha ^{4}$ typo corrected into $\alpha ^{2}$ as should be the case following their procedure). We plot the effect of the PGM spacetime
on the spectroscopic structure of such a KG-Coulomb particle in Figure 7.
We clearly observe that as the PGM parameter decreases from $\alpha =1$ (i.e., Minkowski flat spacetime) the separation between the energy levels increases (which is, in fact, exactly an opposite effect of the PGM/QPGM spacetime of Figure 4) This is obvious from the comparison between equation (\ref{938}) and (\ref{105}).

\section{Shifted KG-oscillators in a PGM spacetime and a WYMM at $\mathcal{F}_{r}\left( r\right) =0$ and $S\left( r\right) =\beta r$}

A Lorentz scalar interaction potential $S\left( r\right) =\beta r$ is commonly used in the study of mass spectroscopy of some quark-antiquark systems (cf. e.g., Gunion and Li \cite{Re028}). When this potential is substituted in Eq.(\ref{e08}) it implies that%
\begin{equation}
\left\{ \tilde{E}^{2}+\partial _{r}^{2}-\frac{\tilde{L}\left( \tilde{L}%
+1\right) }{r^{2}}-\tilde{\beta}^{2}\left( \frac{m_{\circ }}{\beta }%
+r\right) ^{2}\right\} R\left( r\right) =0;\;\tilde{\beta}=\frac{\beta }{%
\alpha }.  \label{111}
\end{equation}%
In this section we shall be interested in the case where%
\begin{equation}
\tilde{L}=0\Longleftrightarrow \sigma ^{2}=\left( eg\right) ^{2}=\ell \left(
\ell +1\right) +2\zeta \left( 1-\alpha ^{2}\right) .  \label{112}
\end{equation}%
This would identify a critical correlation between the Wu-Yang monopole strength $g$ in $\left( eg\right) ^{2}//q\left( \rho \right) \,\rho ^{2}$and
the scalar curvature $R=R_{\nu }^{\nu }=2\left( 1-\alpha ^{2}\right)
/q\left( \rho \right) \,\rho ^{2}$ in $2\zeta \left( 1-\alpha ^{2}\right)
/q\left( \rho \right) \,\rho ^{2}$, so that the central repulsive core $%
\tilde{L}\left( \tilde{L}+1\right) /q\left( \rho \right) \,\rho ^{2}=\tilde{L%
}\left( \tilde{L}+1\right) /r^{2}$ is removed. Which would, in turn, allow the wave functions of the shifted KG-oscillators to be finite and square integrable at the center of the PGM (i.e., at $r=0=\rho $). Under such settings, the Shifted KG-oscillators are described by%
\begin{equation}
\left\{ \tilde{E}^{2}+\partial _{r}^{2}-\tilde{\beta}^{2}\left( \frac{%
m_{\circ }}{\beta }+r\right) ^{2}\right\} R\left( r\right) =0.  \label{113}
\end{equation}%
We may now use the change of variable $z=$ $r+m_{\circ }/\beta $ and recast this equation as%
\begin{equation}
\left\{ \tilde{E}^{2}+\partial _{z}^{2}-\tilde{\beta}^{2}z^{2}\right\}
R\left( z\right) =0,  \label{114}
\end{equation}%
to imply%
\begin{equation}
R\left( r\right) =N\,{}{}\left( \frac{m_{\circ }}{\beta }+r\right) \,\exp
\left( -\frac{\tilde{\beta}\left( \frac{m_{\circ }}{\beta }+r\right) ^{2}}{2}%
\right) {}\,M\left( \frac{3}{4}+\frac{\tilde{E}^{2}}{4\tilde{\beta}},\frac{3%
}{2},\tilde{\beta}\left( \frac{m_{\circ }}{\beta }+r\right) ^{2}\right) .
\label{115}
\end{equation}%
We truncate the the confluent hypergeometric series $M\left(
a,b,z\right) $ into a polynomial of order $n_{r}$ so that%
\begin{equation}
\frac{3}{4}+\frac{\tilde{E}^{2}}{4\tilde{\beta}}=-n_{r},  \label{116}
\end{equation}%
to obtain the shifted KG-oscillators energies as%
\begin{equation}
E=\pm \sqrt{2\alpha \beta \left( 2n_{r}+\frac{3}{2}\right) }  \label{117}
\end{equation}%
and the radial wavefunctions%
\begin{equation}
R\left( r\right) =\tilde{N}\,{}{}\left( m_{\circ }+\beta r\right) \,\exp
\left( -\frac{\alpha \left( m_{\circ }+\beta r\right) ^{2}}{2}\right)
{}M\left( -n_{r},\frac{3}{2},\alpha \left( m_{\circ }+\beta r\right)
^{2}\right) .  \label{118}
\end{equation}

Evidently, at the critical values for the Wu-Yang magnetic monopole strength $\sigma =eg$ given in Eq.(\ref{112}) have made all $\ell $-states to effectively cluster on $\ell =0$ states for a given radial quantum number $%
n_{r}$. That is, all non-zero $\ell $-states have disappear from the spectra as is obvious from Eq.(\ref{117}). For example, for $\ell =0$ states (i.e., $%
S$-states) in (\ref{112}) we have $\sigma ^{2}=\left( eg\right) ^{2}=2\zeta
\left( 1-\alpha ^{2}\right) $ that identifies a critical correlation between the WYMM and the scalar curvature $R$ and/or the PGM parameter $%
\alpha $ through $2\zeta \left( 1-\alpha ^{2}\right) $. For $\ell =1$ states (i.e., $P$-states) we have $\sigma ^{2}=\left( eg\right) ^{2}=2+2\zeta
\left( 1-\alpha ^{2}\right) $ to identify yet another correlation of the
same kind, and so on so forth.

\section{Concluding remarks}

In current study, we have used two correlated metric functions (correlated deformation/transformation functions, if you wish) and transformed a PGM spacetime into a QPGM spacetime, where the WYMM is involved in both cases. Namely, we have studied KG-particles in a PGM/QPGM spacetimes and a WYMM. We have shown that the KG-particles in a QPGM spacetime and a WYMM are isospectral and invariant with KG-particles in the regular PGM spacetime and a WYMM. This is documented in sections 2 and 3. Based on such findings, we use the non-minimal coupling form of the operator $\tilde{D}_{\mu }=D_{\mu }+%
\mathcal{F}_{\mu }$ (where, $\mathcal{F}_{\mu }$ $\in 
%TCIMACRO{\U{211d} }%
%BeginExpansion
\mathbb{R}
%EndExpansion
$, and $D_{\mu }=\partial _{\mu }-ieA_{\mu }$ is the gauge-covariant derivative) of Moshinsky and Szczepaniak \cite{Re0281} and Mirza and Mohadesi \cite{Re0282} \ to discuss KG-oscillators in a QPGM/PGM spacetime and a WYMM and report the effects of both on the spectroscopic structure of the KG-oscillators (documented in Figures 1, 2, and 3). Two sets of correlated dimensionless positive-valued metric functions are used in the process (documented in section 4). Analogously, we show that a Bessel-like functional structure, $\mathcal{F}_{r }$ in Eq. (\ref{935}), would yield KG-Coulomb like particles in a QPGM/PGM spacetime and a WYMM and report/discuss the effects of both monopoles on their spectroscopic structure (as documented in Figures 4, 5, and 6). Moreover, in the PGM spacetime and a WYMM, we discuss a KG-Coulomb and a shifted KG-oscillator as manifestations/byproducts of the Lorentz scalar potentials $S\left( r\right) =\mathcal{K}/r$ and $S\left( r\right) =\mathcal{%
\beta }r$, respectively. 

In connection with the KG-oscillators and KG-Coulomb like particles that are introduced by the non-minimal coupling form of the operator $\tilde{D}_{\mu }=D_{\mu }+%
\mathcal{F}_{\mu }$, in sections 4 and 5, we have observed that as the PGM/QPGM parameter $\alpha$ decreases from $\alpha=1$ value towards $\alpha \sim 0$, the energy levels are destined to cluster on $E=\pm m_{\circ}$.  Whereas, the separations between the energy levels are observed to increase with increasing  WYMM strength $\sigma$ for the KG-oscillators and decrease with increasing  $\sigma$ for the KG-Coulombic particles. However, for the KG-Coulombic particles introduced by the Lorentz scalar potential $S(r)$, we have observed that decreasing the PGM parameter $\alpha$ from $\alpha=1$ would increase the separation between energy levels.

On the PGM spacetime background side, the current methodical proposal suggests that the quantum mechanical systems in the PGM spacetime 

\begin{equation}
ds^{2}=-dt^{2}+\frac{1}{\alpha ^{2}}dr^{2}+r^{2}\left( d\theta ^{2}+\sin
^{2}\theta \,d\varphi ^{2}\right) ,  
\end{equation}%
have infinite number of quantum mechanical systems in the QPGM spacetime%

\begin{equation}
ds^{2}=-dt^{2}+\frac{f\left( \rho \right) }{\alpha ^{2}}d\rho ^{2}+q\left(
\rho \right) \,\rho ^{2}\left[ d\theta ^{2}+\sin ^{2}\theta \,d\varphi ^{2}%
\right] ,  
\end{equation}%
that share the same energy levels,  provided that the two metric functions $f( \rho )$ and $q( \rho)$ are positive- valued dimensionless correlated (through Eq. (\ref{e02}) ) multipliers.  We have, hereinabove, tested  this fact for scalar KG -particles. In a recent work \cite{Re29}, we have also tested this for non-relativistic Schr\"{o}dinger particles in PGM/QPGM spacetime and have arrived at the same conclusion. It would be, therefore, interesting to investigate this fact for Dirac-particles (Dirac-oscillators in particular). Such investigations should also include other quantum mechanical phenomena like scattering, thermodynamical effects, etc.


\begin{thebibliography}{99}
\bibitem{Re1} T W B Kibble, Phys. Rep. \textbf{67} (1980) 183.

\bibitem{Re2} A. Vilenkin, Phys. Rep. \textbf{121} (1985) 263.

\bibitem{Re021} A. Vilenkin, Phys. Rev. D \textbf{23} (1981) 852.

\bibitem{Re0211} M. O Katanaev, I. V. Volovich, Ann. Phys. \textbf{216}
(1992).

\bibitem{Re0212} R. A. Puntigam, H. H. Soleng, Class. Quant. Gravit. \textbf{%
14} (1997) 1129.

\bibitem{Re3} A. Vilenkin, Phys. Rev. Lett. \textbf{46} (1988) 1169.

\bibitem{Re4} A. Vilenkin, Phys. Lett. B \textbf{133} (1983) 177.

\bibitem{Re5} W A Hiscock, Phys. Rev. D \textbf{31} (1985) 3288.

\bibitem{Re6} B Linet, Gen. Relativ. Gravit. \textbf{17} (1985) 1109,

\bibitem{Re7} M. Barriola, A. Vilenkin, Phys. Rev. Lett. \textbf{63} (1989)
341.

\bibitem{Re8} A L Cavalcanti de Oliveira, E R Bezerra de Mello, Class Quant.
Grav. \textbf{23} (2006) 5249.

\bibitem{Re81} T.R.P. Caram\^{e}s, J.C. Fabris, E R Bezerra de Mello, H
Belich, Eur. Phys. J. C \textbf{77} (2017) 496.

\bibitem{Re9} R L L Vit\'{o}ria, H Belich, Phys. Scr. \textbf{94} (2019)
125301.

\bibitem{Re10} C Furtado, F Moraes, J. Phys. A: Math. Gen. \textbf{33}
(2000) 5513.

\bibitem{Re101} T T Wu, C N Yang, Nucl. Phys. B \textbf{107} (1976) 365.

\bibitem{Re1011} T T Wu, C N Yang, Phys. Rev. D \textbf{12} (1975) 3845.

\bibitem{Re102} E R Bezerra de Mello, Class. Quant. Grav. \textbf{19} (2002)
5141.

\bibitem{Re103} J Spinelly, U de Freitas, E R Bezerra de Mello, Phys. Rev. D 
\textbf{66} (2002) 024018.

\bibitem{Re11} E A F Bragan\c{c}a, R L L Vit\'{o}ria, H Belich, E R Bezerra
de Mello, Eur. Phys. J. C \textbf{80} (2020) 206.

\bibitem{Re12} M Moshinsky, A Szczepaniak, J. Phys. A: math. Gen. \textbf{22}
(1989) L817.

\bibitem{Re121} B Mirza, M Mohandesi, Commun. Theor. Phys. \textbf{42}
(2004) 664.

\bibitem{Re13} K Bakke, H F Mota, Eur. Phys. J. Plus \textbf{133} (2018) 409.

\bibitem{Re131} A Boumali, H Aounallah, Adv. High Energy Phys. \textbf{2018}
(2018) 1031763.

\bibitem{Re132} G A Marques, E R Bezerra de Mello, Class. Quant. Gravit. 
\textbf{19} (2002) 985.

\bibitem{Re1321} S S Alves, M M Cunha, H Hassaabadi, E O Silva, Universe 
\textbf{9} (2023) 132.

\bibitem{Re133} E R Bezerra de Mello, C Furtado, Phys. Rev. D \textbf{56}
(1997) 1345.

\bibitem{Re1331} F. Ahmed, Phys. Scr. \textbf{98 }(2023) 015403.

\bibitem{Re1332} F. Ahmed, Sci. Rep. \textbf{12} (2022) 8794.

\bibitem{Re14} J Cravalho, C Furtado, F Moraes, Phys. Rev. A \textbf{84}
(2011) 032109.

\bibitem{Re15} N A Rao, B A Kagali, Mol. Phys. Lett. A \textbf{19} (2004)
2147.

\bibitem{Re16} K Bakke, C Furtado, Ann. Phys. \textbf{336} (2013) 489.

\bibitem{Re17} P Strange, L H Ryder, Phys. Lett. A \textbf{380} (2016) 3465.

\bibitem{Re18} O Mustafa, Ann. Phys. \textbf{440} (2022) 168857.

\bibitem{Re19} O Mustafa, Eur. Phys. J. C \textbf{82} (2022) 82.

\bibitem{Re20} O Mustafa, Ann. Phys. \textbf{446} (2022) 169124.

\bibitem{Re21} O Mustafa, Eur. Phys. J. Plus \textbf{138} (2023) 21.

\bibitem{Re211} O. Mustafa, Phys. Lett. B \textbf{839} (2023) 137793.

\bibitem{Re22} A Boumali, N Messai, Can. J. Phys. \textbf{92} (2014) 11.

\bibitem{Re23} K Bakke, C Furtado, Ann. Phys. \textbf{355} (2015) 48.

\bibitem{Re24} R L LVit\'{o}ria, H Belich, Eur. Phys. J. C \textbf{78}
(2018) 999.

\bibitem{Re25} R L LVit\'{o}ria, K Bakke, Eur. Phys. J. Plus \textbf{133}
(2018) 490.

\bibitem{Re26} R L LVit\'{o}ria, H Belich, K Bakke, Eur. Phys. J. Plus 
\textbf{132} (2017) 25.

\bibitem{Re27} J Cravalho, A M Cravalho, E Cavalcante, C Furtado, Eur. Phys.
J. C \textbf{76} (2016) 365.

\bibitem{Re271} H. Hassanabadi, M. Hosseinpour, Eur. Phys. J. C \textbf{76}
(2016) 553.

\bibitem{Re272} H. Hassanabadi, S Sargolzaeeipor, B H Yazarloo, Few-Body
syst. \textbf{56} (2015) 115.

\bibitem{Re028} J. F. Gunion, L. F. Li, Phys. Rev. D \textbf{12 (}1975) 3583.

\bibitem{Re0281} M. Moshinsky, A. Szczepaniak, J. Phys. \textbf{A}: Math.
Gen. \textbf{22} (1989) L817.

\bibitem{Re0282} B. Mirza, M. Mohadesi, Commun. Theor. Phys. \textbf{42}
(2004) 664.

\bibitem{Re29} O. Mustafa: arXiv: 2308.10176 , " Schr\"{o}dinger oscillators in a deformed point-like global monopole spacetime and a Wu-Yang magnetic monopole: position-dependent mass correspondence and isospectrality".

\end{thebibliography}
\end{document}